\documentclass[preprint,12pt]{elsarticle}

\usepackage{amssymb}

\usepackage{epstopdf}
\usepackage{url}
\usepackage[usenames,dvipsnames]{color}
\usepackage{listings}

\definecolor{MyDarkGreen}{rgb}{0.0,0.4,0.0}

\journal{Journal of Computational Physics}

\begin{document}

\begin{frontmatter}


 \title{Title\tnoteref{label1}}

\title{The Energy Conserving Particle-in-Cell Method}


\author{Stefano Markidis and Giovanni Lapenta}

\address{Centre for Plasma Astrophysics, Katholieke Universiteit Leuven, Celestijnenlaan 200B, B-3001 Leuven, Belgium}

\begin{abstract}
A new Particle-in-Cell (PIC) method, that conserves energy exactly, is presented. The particle equations of motion and the Maxwell's equations are differenced implicitly in time by the midpoint rule and solved concurrently by a Jacobian-free Newton Krylov (JFNK) solver. 
Several tests show that the finite grid instability is eliminated in energy conserving PIC simulations, and the method correctly describes the two-stream and Weibel instabilities, conserving exactly the total energy. The computational time of the energy conserving PIC method increases linearly with the number of particles, and it is rather insensitive to the number of grid points and time step. The kinetic enslavement technique can be effectively used to reduce the problem matrix size and the number of JFNK solver iterations.

\end{abstract}


\end{frontmatter}


\section{Introduction}
The Particle-in-Cell (PIC) method is one of the most used numerical methods for the solution of the collision-less kinetic equation of plasmas. The majority of PIC schemes has the property of conserving exactly the system total momentum~\cite{birdsall,hockney}, while it does not conserve the system total energy. In fact typically PIC methods, that use explicit differentiation in time (explicit PIC schemes), tend to increase the total energy of the system by numerical heating~\cite{birdsall,hockney}, while PIC methods, that use implicit differentiation in time (implicit PIC schemes), tend to decrease the total energy of the system by numerical cooling~\cite{brackbill-cohen-85}. In the study of plasma physics instabilities, where one kind of energy is converted to another one, it is important to ensure that energy is not created spuriously by the numerical scheme in use. In fact, the numerical heating introduces spurious energy that can feed erroneously plasma instabilities, leading to unphysical results.

A new class of PIC methods were developed at beginning of the Seventies to address the problem of the total energy conservation. An "energy conserving" PIC method was first proposed by Lewis in 1970~\cite{Lewis}. However, the Lewis' "energy conserving" scheme conserves energy only in the limit of zero time steps, while accuracy errors preclude the possibility of having exact energy conservation with finite time steps~\cite{birdsall,hockney}. The scheme does not lead to the exact energy conservation, but instead to an improved energy conservation at a given time step if compared to momentum conserving PIC methods. The "energy conserving" method was derived from the Lagrangian formalism, and proved to be more robust against aliasing instabilities, such as the finite grid instability. An in-depth analysis of the "energy conserving" PIC method is presented by Langdon in Ref.~\cite{EClangdon} and in textbooks~\cite{birdsall,hockney}. 

Differently from previous "energy conserving" PIC methods, the proposed PIC scheme has the property of conserving exactly the total energy not only in the limit of zero time step, but also with finite time step. The new PIC method conserves the system total energy with a precision that depends only on the error tolerance value of the iterative solver in use. The scheme is not derived from the Lagrangian formalism but instead from the Newton approach. Energy conservation is achieved by using the midpoint integration rule for particle and field equations, and a convenient discretization of the discrete spatial operators. 

The energy conserving PIC method requires the concurrent solution of the equation of motion for each particle and of the field equations for the electric and magnetic fields at each grid point, posing two challenges. First, the direct solution of the implicit numerical equations of PIC method implies the computation of a non-linear system. Non-linearity arises from the coupling between particles and fields variables through the interpolation functions of the PIC method. In the proposed PIC scheme, the non-linear equations are solved by a Newton Krylov solver~\cite{Kelley:2003,Knoll:2004}. Despite the belief that the iterative solution of such equations could hardly converge~\cite{langdon:1979}, it has been proven that such PIC methods, that are called {\em fully implicit}, are convergent~\cite{markidisPHDthesis, Kim2005}. The second challenge is that the energy conserving PIC scheme requires the solution of a very large matrix whose rank is of the order of the number of particles (the number of particles is considerably higher than the number of grid points in typical PIC simulations). For this reason, implementations of fully implicit PIC methods are based on the matrix-free Jacobian-free solvers to avoid the storage of  the matrix, and Jacobian coefficients~\citep{markidisPHDthesis}. Previous implementations of fully implicit PIC methods, as those presented in Refs.~\cite{markidisPHDthesis, Kim2005}, were limited to electrostatic simulation and more importantly their formulation does not imply the total energy conservation. The new energy conserving PIC method is still based on the solution of coupled non-linear equations by a Jacobian-free Newton Krylov (JFNK) solver, but on the contrary it is formulated for the electromagnetic case and conserves exactly the total energy.

This paper presents the algorithm, the properties, the implementation, the simulation and performance results of the energy conserving PIC method. It is organized as follows. Section 2 introduces the governing equations, shows their discretization in time and space and explains in detail the numerical algorithm. Sections 3,4,5 analyze the properties of the proposed method: the energy and momentum conservation, and the numerical stability. The implementation of an energy conserving PIC code is discussed in Section 6. Section 7 presents first the results of a Maxwellian plasma simulation, that is robust against the finite grid instability, and then of the two-stream and Weibel instabilities. The computational performance of the proposed method, and a technique to reduce the problem matrix size are shown in Section 8. Finally, Section 9 concludes the paper summarizing the algorithm and its properties. In Appendices A and B, a skeleton version of the energy conserving PIC code in Matlab/Octave programming language is provided.

\section{Algorithm}
In the PIC method $N_s$ computational particles of the different $n_s$ species with label $s$ mimic the real behavior of electrons and ions~\cite{dawson1983}. Each computational particle is characterized by a position $\mathbf{x}_p$ and a velocity $\mathbf{v}_p$, whose evolution is described by the equation of motion (here and thereafter in CGS units):
\begin{equation}
\label{motion}
\left\{
\begin{array}{l}
{d {\bf x}_p}/{dt} ={\bf v}_p \\
{d {\bf v}_p}/{dt} =q_s/m_s\left( {\bf E}_p + {\bf v}_p/c \times {\bf B}_p\right),
\end{array}
\right.
\end{equation}
where $q_s/m_s$ are the charge to mass ratio of the species $s$. ${\bf E}_p $, and ${\bf B}_p $ are the electric and magnetic fields acting on the particle $p$ and they are calculated by interpolation from ${\bf E}_g$ and ${\bf B}_g $, the values of the electric and magnetic field on the $N_g$ grid points, through the use of the interpolation function $W({\bf x}_g-{\bf x}_p) $:
\begin{equation}
{\bf E}_p=\sum_g^{N_g} {\bf E}_g W({\bf x}_g-{\bf x}_p) \quad \quad {\bf B}_p=\sum_g^{N_g} {\bf B}_g W({\bf x}_g-{\bf x}_p) .
\label{interp}
\end{equation}
The Cloud-in-Cell interpolation functions ~\cite{birdsall,hockney} are used:
\begin{equation}
 \label{interpW}
W({\bf x}_g-{\bf x}_p) =
\left\{
\begin{array}{l}
 1 - |{\bf x}_g-{\bf x}_p|/\Delta x \quad \textup{if} \quad |{\bf x}_g-{\bf x}_p| < \Delta x \\
0  \quad \textup{otherwise}  .
\end{array}
\right.
\end{equation}
Equations \ref{motion} are differenced in time using the implicit midpoint integration rule \cite{Hairer:2002}:
\begin{equation}
\label{dif_eom}
\left\{
\begin{array}{l}
\mathbf{v}_p^{n+1} = \mathbf{v}_p^{n} + q_s /m_s \Delta t (\mathbf{\bar{E}}_p +  \mathbf{\bar{v}}_p/c \times \mathbf{\bar{B}}_p)  \\
\mathbf{x}_p^{n+1} = \mathbf{x}_p^{n} + \mathbf{\bar{v}}_p\Delta t
\end{array} 
\right. ,
\end{equation}
where $n$ is the time level and the bar variables are the average in time of the quantities, and they are defined as:
\begin{eqnarray}
\mathbf{\bar{x}}_p = (\mathbf{x}_p^{n+1} + \mathbf{x}_p^{n})/2 \quad \quad \quad \mathbf{\bar{v}}_p = (\mathbf{v}_p^{n+1} + \mathbf{v}_p^{n})/2\\
\label{EBmediop}
\mathbf{\bar{E}}_p = \sum_g^{N_g} \mathbf{\bar{E}}_g W(\mathbf{x}_g - \mathbf{\bar{x}}_p)\quad \quad \quad \mathbf{\bar{B}}_p = \sum_g^{N_g} \mathbf{\bar{B}}_g W(\mathbf{x}_g - \mathbf{\bar{x}}_p)  \\
\label{EBmedio}
\mathbf{\bar{E}}_g = (\mathbf{E}_g^{n+1} + \mathbf{E}_g^{n})/2 \quad \quad \quad \mathbf{\bar{B}}_g = (\mathbf{B}_g^{n+1} + \mathbf{B}_g^{n})/2 .
\end{eqnarray}
It is possible to rewrite Equations \ref{dif_eom} in terms of $\mathbf{\bar{v}}_p$ after a series of algebraic manipulations~\citep{brackbill-cohen-85,markidisPHDthesis}:
\begin{eqnarray}
\label{vhat2}
\tilde{\mathbf{v}}_p&=&\mathbf{v}_p^n+\frac{q_s\Delta t}{2m_s}\mathbf{\bar{E}}_p\\
\label{vn+1/2}
\mathbf{\bar{v}}_p&=&\frac{\tilde{\mathbf{v}}_p+\frac{q_s\Delta
t}{2m_s c}\bigl(\tilde{\mathbf{v}}_p\times\mathbf{\bar{B}}_p+\frac{q_s\Delta
t}{2m_s c}(\tilde{\mathbf{v}}_p\cdot\mathbf{\bar{B}}_p)\mathbf{\bar{B}}_p\bigr)}{(1+\frac{q_s^2\Delta t^2}{4m_s^2c^2}{\bar{B}_p}^2)},
\end{eqnarray}
and the equation of motion becomes:
\begin{equation}
\label{dif_eom2}
\left\{
\begin{array}{l}
\mathbf{v}_p^{n+1} = 2 \mathbf{\bar{v}}_p - \mathbf{v}_p^{n}  \\
\mathbf{x}_p^{n+1} = \mathbf{x}_p^{n} + \mathbf{\bar{v}}_p\Delta t .
\end{array} 
\right. 
\end{equation}
The evolution of the electric and magnetic fields is determined by solving the Maxwell's equations:
 \begin{equation}
 \label{Maxwell}
\left\{
\begin{array}{l}
\nabla \cdot \mathbf{E} =  4 \pi  \rho \\
\nabla \cdot \mathbf{B} = 0 \\
 {1}/{c}\,{\partial \mathbf{E}}/{\partial t}  = \nabla \times \mathbf{B}  - {4 \pi}/{c}\, \mathbf{J}  \\
 {1}/{c}\,{\partial \mathbf{B}}/{\partial t}  = - \nabla \times \mathbf{E},
\end{array}
\right.
\end{equation}
The implicit midpoint scheme is used to discretize the Maxwell's equations. The Faraday's and Ampere's laws are differenced in time as follows:
\begin{equation}
\label{MaxwDiscrete}
\left\{
\begin{array}{l}
\mathbf{E}_g^{n+1} - \mathbf{E}_g^{n} = c \nabla \times \mathbf{\bar{B}}_g \Delta t - 4 \pi \mathbf{\bar{J}_g} \Delta t = c/2 \nabla \times (\mathbf{B}_g^{n+1}+\mathbf{B}_g^{n}) \Delta t - 4 \pi \mathbf{\bar{J}_g}  \\
\mathbf{B}_g^{n+1} - \mathbf{B}_g^{n} = - c \nabla \times \mathbf{\bar{E}}_g \Delta t = - c/2 \nabla \times (\mathbf{E}_g^{n+1}+\mathbf{E}_g^{n}) \Delta t.
\end{array}  
\right. 
\end{equation}
The average current density $\mathbf{\bar{J}_g} $ is calculated from the particle average positions and velocities by interpolation:
\begin{equation}
\label{Jmedio}
\mathbf{\bar{J}}_g = \sum^{n_s}_s \sum^{N_s}_p q_s \mathbf{\bar{v}}_p W(\mathbf{x}_g - \mathbf{\bar{x}}_p)/V_g ,
\end{equation}
where $V_g$ is the volume of cell $g$.

The discretization of spatial operators must be chosen carefully to ensure that the vector identities, that are valid in the continuous space, hold on the discrete grid also. To achieve this, the Yee's lattice~\citep{Mimetic,Yee:1966} discretization of the spatial operators in Equations \ref{MaxwDiscrete} is used. Taking a uniform rectangular grid for simplicity, the different components of the electromagnetic field and of the current densities are calculated on the cell center (half integer index) and on the cell vertices (integer index) according to the Yee's lattice configuration:
\begin{equation}
\begin{array}{l}
\mathbf{E}_g = (E^x_{i,j+1/2,k+1/2},E^y_{i+1/2,j,k},E^z_{i+1/2,j,k}) \\
\mathbf{B}_g = (B^x_{i+1/2,j,k},B^y_{i,j+1/2,k+1/2},B^z_{i,,j+1/2,k+1/2}) \\
\mathbf{\bar{J}}_g = (\bar{J}^x_{i,j+1/2,k+1/2},\bar{J}^y_{i+1/2,j,k+1/2},\bar{J}^z_{i+1/2,j+1/2,k}) .
\end{array} 
\end{equation}
The discrete operator $\nabla$  is a centered difference in space (second order accurate):
\begin{equation}
\begin{array}{l}
\nabla \mathbf{f}_{i,j,k} = ((f_{i+1/2,j,k}-f_{i-1/2,j,k})/\Delta x,(f_{i,j+1/2,k}-f_{i,j-1/2,k})/\Delta y, \\ 
(f_{i,j,k+1/2}-f_{i,j,k-1/2})/\Delta z)  .
\end{array} 
\end{equation}
The properties hold for the discrete operator $\nabla$ in the chosen spatial discretization:
\begin{equation}
\label{vect_ident}
\nabla \cdot \nabla \times \mathbf{f} = 0 \quad \quad \nabla \cdot (\mathbf{f} \times \mathbf{h}) = \mathbf{h} \cdot (\nabla \times \mathbf{f}) -  \mathbf{f} \cdot (\nabla \times \mathbf{h})
\end{equation}

\begin{figure}
\centering
\includegraphics[width=\columnwidth]{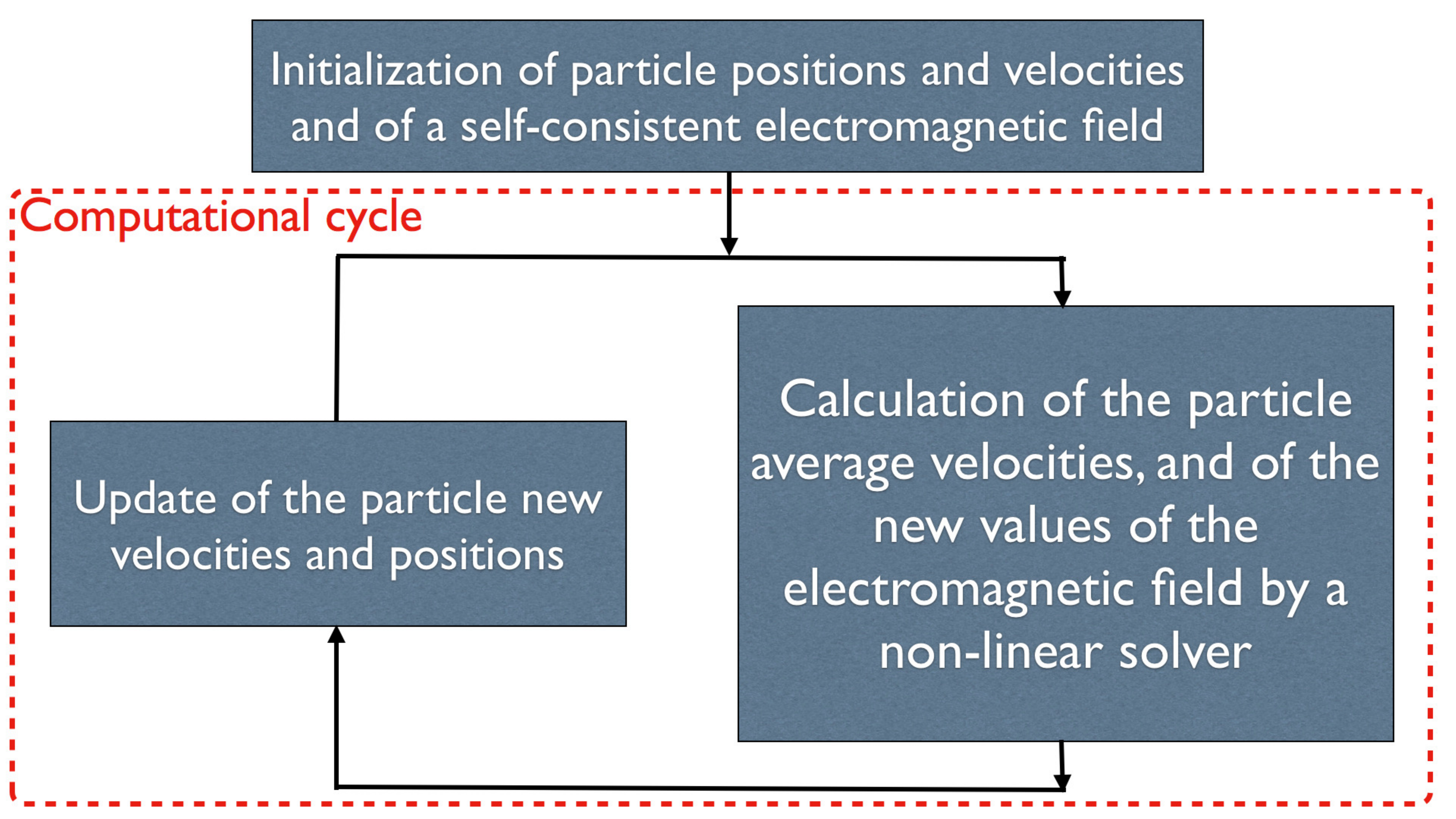}
\caption[]
{ \label{fig:ECPICalgorithm}
The energy conserving PIC algorithm. After the simulation has been initialized, the computational cycle (non-linear solver stage, and particle update) is repeated.}  
\end{figure}

The energy conserving PIC method is based on the concurrent solution of the coupled Equations \ref{vn+1/2} and \ref{MaxwDiscrete} by a non-linear solver. The algorithm is summarized in Figure \ref{fig:ECPICalgorithm}.
The simulation is initialized first, setting the particles positions and electromagnetic field self consistently. At each PIC computational cycle, Equations \ref{vn+1/2} and \ref{MaxwDiscrete} (non-linearly coupled by Equation \ref{Jmedio}) are solved by a JFNK solver. The governing equations have been time differenced with the implicit midpoint method, and solved in terms of $\mathbf{\bar{v}}_p$, $\mathbf{E}^{n+1}$, and $\mathbf{B}^{n+1}$. The spatial operators of Equations \ref{MaxwDiscrete} are differenced in space on the Yee's lattice. Once $\mathbf{\bar{v}}_p$ has been calculated with the solver, the new particle positions and velocities are simply updated with Equation \ref{dif_eom2}. 

\subsection{The Electrostatic Limit}
The electrostatic formulation of the energy conserving PIC method can be derived by considering an unmagnetized plasma. In this case, the particle average velocity Equation \ref{vn+1/2} simply reduces to:
\begin{equation}
\label{vnES}
\mathbf{\bar{v}}_p = \mathbf{v}^n_p + \frac{q_s}{2 m_s} \mathbf{\bar{E}}_p \Delta t = \mathbf{v}^n_p + \frac{q_s}{4 m_s} (\mathbf{E}^{n+1}_p (\mathbf{\bar{x}}_p) + \mathbf{E}^{n}_p(\mathbf{\bar{x}}_p))  \Delta t  .
\end{equation}
The evolution of the electric field is determined by the Ampere's law:
\begin{equation}
\label{MaxwDiscreteES}
\mathbf{E}_g^{n+1} - \mathbf{E}_g^{n} = - 4 \pi \mathbf{\bar{J}_g} \Delta t .
\end{equation}

\subsection{Divergence Equations}
\begin{figure}
\centering
\includegraphics[width=\columnwidth]{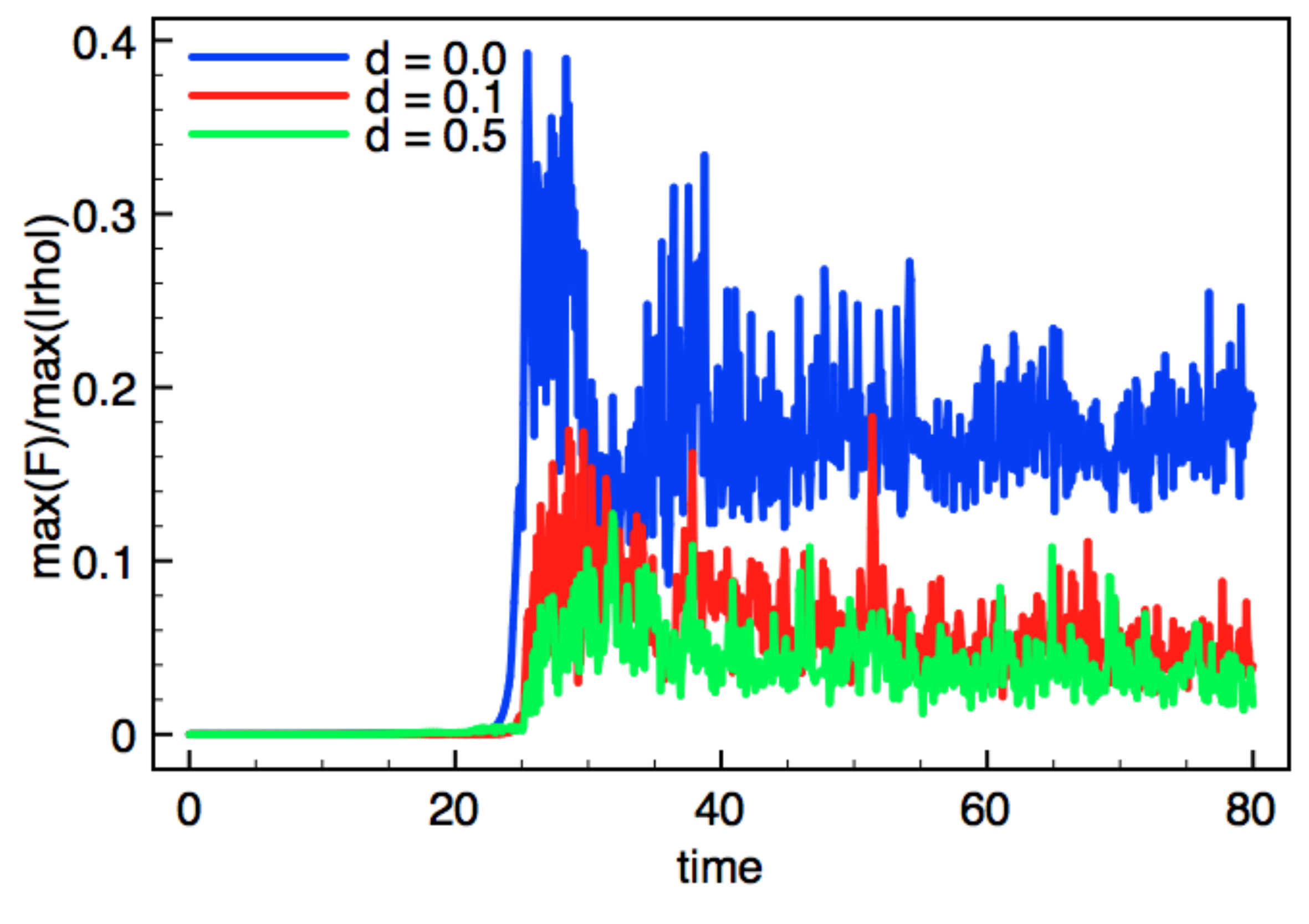}
\caption[]
{ \label{Fig:GaussError}
Maximum violation of Gauss' law divided by the maximum charge density in a two-stream instability simulation. The blue line represents the numerical error without the pseudo current correction, while the red and green lines show its evolution with a pseudo current correction with $d=0.1c^2/\omega_{pe}$ and $0.5c^2/\omega_{pe}$ respectively.}
\end{figure}

Only the Faraday's and Ampere's equations were considered so far, while the two divergence equations of the system \ref{Maxwell} were not taken in account. It easy to show that the equation $\nabla \cdot \mathbf{B} = 0$ is always satisfied if it is initially~\cite{birdsall}. 
Moreover, the Gauss' law $\nabla\cdot \mathbf{E} = 4 \pi \rho$ equation is automatically satisfied by Equation \ref{MaxwDiscrete}, if the charge continuity equation $\partial \rho /\partial t + \nabla \cdot \mathbf{J} = 0$ holds true. In Particle-in-Cell methods, the charge continuity equation is not always satisfied because discrepancies between the interpolation of charge and current densities into the grid~\cite{birdsall}. In fact, the definition of the current density in the energy conserving PIC method (Equation \ref{Jmedio}) does not satisfy the charge density continuity equation. In this case, the method is said not to conserve the charge. As pointed out in Ref.\cite{Marder:1987}, the Gauss' law can be regarded as a conservation principle: it is not a strictly necessary equation for describing the evolution of electromagnetic fields, but its violation introduces numerical errors in the simulation and might lead to unphysical behavior of the simulated plasma. All the energy conserving PIC simulations, based on the non conservative current density definition, have been first initialized solving the Gauss' law, ensuring there is no error due to the violation of the charge continuity equation initially. Then, the charge conservation has been constantly checked. The error did not grow considerably or lead to unphysical behavior of the simulated plasma. The numerical error can be reduced using the pseudo current method as in Refs.\cite{Marder:1987,Langdon:1992}. In this approach, $F^n_g$ is defined as the violation of the Gauss' law in the cell $g$ at the time level $n$, $F_g^n = \nabla \cdot \mathbf{E}_g^n - 4 \pi \rho^n_g$, and its gradient added to the Ampere's equation:
\begin{equation}
\label{PseudoCurrent}
\frac{\mathbf{E}_g^{n+1} - \mathbf{E}_g^{n}}{\Delta t} = c \nabla \times \mathbf{\bar{B}}_g - 4 \pi \mathbf{\bar{J}_g} +   4 \pi d \nabla \bar{F_g}, 
\end{equation}
where $d$ is a parameter that regulates the charge conservation, and $\bar{F_g}$ is $1/2(F^{n+1}_g+F^{n}_g)$ (differently from Ref.\cite{Marder:1987}, F is calculated at $n+1/2$ and solved implicitly). Taking the divergence of Equation \ref{PseudoCurrent}:
\begin{equation}
\frac{F^{n+1}_g - F^{n}_g}{d \Delta t} - \nabla^2 \bar{F_g} = - (\frac{\rho^{n+1}_g+\rho^{n}_g}{\Delta t} + \nabla \cdot  \mathbf{\bar{J}_g})
\end{equation}
The left side of the equation above is the heat equation. If $F_g$ is fixed to zero at the boundaries initially and at each simulation time step, the error due to non conservation of charge diffuses away with a rate determined by the parameter $d$. Figure \ref{Fig:GaussError} shows the error due to the violation of Gauss' law using $d$ equal to 0 (no pseudo current correction), and to $0.1 c^2/\omega_{pe}$, $0.5 c^2/\omega_{pe}$ in a simulation of the two-stream instability. It has been found that the pseudo current correction decreases the error related to the non conservation of charge. No major differences appeared in runs with and without the pseudo current correction: the plasma instabilities under study started at the same time and presented the same growth rate. It is clear from Figure \ref{Fig:GaussError} that the numerical error builds up slowly in the case of no pseudo current correction (blue line).

\section{Energy Conservation}
The discretized equations of the proposed PIC method have the property of conserving exactly the total energy. In fact, the variation of the magnetic and electric fields energies ($W_E$, $W_B$) during the time step $\Delta t$ is:
\begin{equation}
\Delta (W_E + W_B)  = \frac{1}{8\pi} \sum_g^{N_g} \bigg( (\mathbf{E}_g^{n+1})^2 -  (\mathbf{E}_g^{n})^2 + (\mathbf{B}_g^{n+1})^2 -  (\mathbf{B}_g^{n})^2 \bigg) V_g
\end{equation}
Substituting Equations \ref{MaxwDiscrete} in the formula above: 
\begin{equation}
\label{varE}
\begin{array}{l}
\Delta (W_E + W_B)  =  \frac{1}{4\pi} \sum_g^{N_g} \bigg( \mathbf{\bar{E}}_g \cdot (c \nabla \times \mathbf{\bar{B}}_g - 4\pi \mathbf{\bar{J}}_g ) -    \mathbf{\bar{B}}_g \cdot (c \nabla \times \mathbf{\bar{E}}_g \Delta t)\bigg) \Delta t V_g\\
    =   \sum_g^{N_g}  \bigg(- \mathbf{\bar{E}}_g \cdot \mathbf{\bar{J}}_g - \nabla \cdot \mathbf{\bar{S}}_g \bigg) \Delta t V_g.
\end{array}
\end{equation}
The vector identities \ref{vect_ident}, that hold in the chosen spatial discretization, have been used. The first term of the equation above represents the work of the field on the particles, while the second term $\mathbf{\bar{S}}_g = c/4 \pi (\mathbf{\bar{E}}_g \times \mathbf{\bar{B}}_g )$ is the average Poynting flux. Its contribution to the energy variation is zero in an isolated system. If the expression for the average current density $\mathbf{\bar{J}}_g $ (Equation \ref{Jmedio}) is substituted in the formula above, then:
\begin{equation}
\begin{array}{l}
\Delta (W_E + W_B)  =   \sum_g^{N_g}  \bigg(- \mathbf{\bar{E}}_g \cdot (\sum^{n_s}_s \sum^{N_s}_p q_s \mathbf{\bar{v}}_p W(\mathbf{x}_g - \mathbf{\bar{x}}_p)/V_g) \bigg) \Delta t V_g\\
= -\sum^{n_s}_s \sum^{N_s}_p q_s \Delta t \mathbf{\bar{v}}_p \cdot \sum_g^{N_g} \mathbf{\bar{E}}_g W(\mathbf{x}_g - \mathbf{\bar{x}}_p)    \\
= - \sum^{n_s}_s \sum^{N_s}_p m_s \mathbf{\bar{v}}_p \cdot (\mathbf{v}_p^{n+1} - \mathbf{v}_p^{n+1} - \sum_g^{N_g} \mathbf{\bar{v}}_p/c \times \mathbf{\bar{B}}_g)   \\
=  - \sum^{n_s}_s \sum^{N_s}_p 1/2 m_s((\mathbf{v}_p^{n+1})^2  - (\mathbf{v}_p^{n})^2 ) .
\end{array}
\end{equation}
Thus the numerical scheme presented in Section 2 conserves the total energy. It must be pointed out that a different definition of $\mathbf{\bar{v}}_p$ in Eq. \ref{vn+1/2}, using $\mathbf{B}_p^n$ instead of $\mathbf{\bar{B}}_p$, still leads to the conservation energy because the magnetic field does no work on a charged particle. In addition, the conservation of energy is a consequence of the definition of the current density, Equation \ref{Jmedio}, and PIC schemes with different techniques for the computation of current density to ensure charge conservation do not conserve the total energy. 
\section{Momentum Conservation}
\begin{figure}
\centering
\includegraphics[width=\columnwidth]{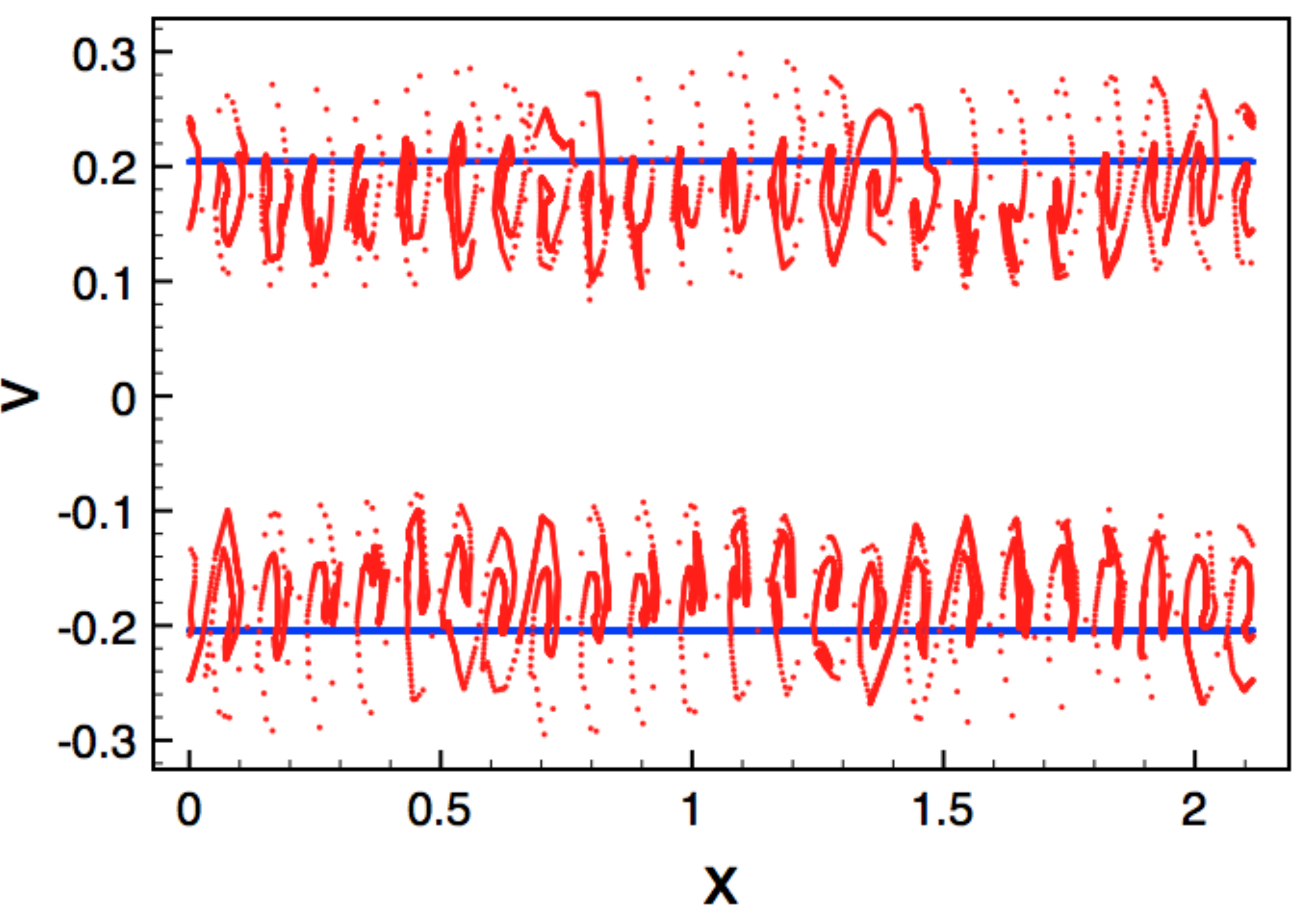}
\caption[]
{ \label{Fig:NonConserMomentum}
Phase space of simulation of two cold counter-streaming electron beams with the energy conserving PIC method. The blue dots represent particles in the initial configuration, while the red dots show the phase space at $t = 13\;\omega_{pe}^{-1}$. An instability due to the non conservation of momentum is visible as red vortices in the phase space.}
\end{figure}

The energy conserving PIC method does not conserve momentum, as reported by Langdon in Ref.\cite{EClangdon} for other "energy conserving" PIC schemes.The non conservation of momentum in energy conserving PIC schemes is due to spurious particle self-forces arising from the non smoothness of the current density deposition to the grid. The relevance of the particle self-forces depends largely on the interpolation functions in use, on the number of particles per cell, and on the grid spacing~\cite{EClangdon}. Self-forces might trigger a macroscopic instability, as shown in Ref.\cite{EClangdon}. For instance in Figure \ref{Fig:NonConserMomentum}, two cold electron beams, composed by 10000 particles (in blue in the Figure \ref{Fig:NonConserMomentum}) are moving initially at opposite velocities $\pm 0.2c$ in a one dimensional 2.053 $c/\omega_{pe}$ long periodic domain with simulation time step $\Delta t = 0.5$ and 64 grid cells. The simulation is completed with the energy conserving PIC method. An aliasing instability develops phase space vortices (red dots in Figure \ref{Fig:NonConserMomentum}) later in time.
A parametric study has been completed varying the grid spacing, the time step and $v_c$, the characteristic velocity of the plasma (e.g. the thermal velocity in a Maxwellian plasma or the drift velocity in beams). It has been found empirically that this aliasing instability disappears if the condition 
\begin{equation}
\label{MomentumCon}
v_c \Delta t /\Delta x < 1.5
\end{equation}
is satisfied. Thus, a modified Courant-Friederics condition, that restricts the maximum particle motion to one and half cell per time step, must be satisfied. The velocity $v_c$ can be connected to the fast electron scales, but no significant limitation arises when one desires the simulation of ion dynamics. It is important to note that, given a $v_c$ that is related to the electron scales (e.g., electron thermal velocity), the ratio $ \Delta t /\Delta x$ can be still adjusted to satisfy Eq. \ref{MomentumCon}, to follow the slower ion scales. In fact, such numerical condition does not preclude the possibility of simulations with large $\Delta t$, but it only requires that the grid spacing is chosen accordingly to satisfy the condition \ref{MomentumCon}. It must be pointed out that this constraint, that arises from the non conservation of momentum, is very similar to the one of the implicit moment PIC method \cite{brackbill-cohen-85} and for this reason the energy conserving PIC method allows simulation time steps that are as large as the one allowed by the implicit moment PIC method.

\section{Numerical Stability}
The numerical stability of PIC methods can be determined by studying the plasma numerical dispersion relation~\cite{lindman:1970}, following the examples of Langdon in Ref.\cite{langdon:1979}, and of Brackbill and Forslund in Ref.\cite{brackbill-cohen-85} for the electrostatic limit. In this approach, the particle equation of motion is linearized, and the electric field is assumed to have an $\exp(i\omega t)$ dependence. The linearized equation of motion is Fourier transformed in $x$, and then the perturbed charge density and the plasma susceptibility are calculated. Following this approach, the numerical dispersion of the energy conserving PIC method results: 
\begin{equation}
\label{disp_implicit}
1- (\frac{\omega_{pe} \Delta t}{2})^2 \int_{-\infty}^{+\infty} f_0(\mathbf{v}) \frac{\cos((\omega - \mathbf{k} \cdot \mathbf{v})\frac{\Delta t}{2})}{\sin^2((\omega - \mathbf{k} \cdot \mathbf{v}) \frac{\Delta t}{2})} d\mathbf{v} = 0,
\end{equation}
where $f_0(\mathbf{v})$ is the equilibrium distribution function, $\omega_{pe} = \sqrt{4 \pi n_eq_e^2/m_e}$ is the plasma frequency, $n_e$ is the plasma density, and $\mathbf{k}$ is the wave vector. In the case of cold plasma with $f_0(\mathbf{v}) = \delta(\mathbf{v})$, the numerical dispersion relation reduces to:
\begin{equation}
\label{disp_beam2}
\tan(\frac{\omega \Delta t}{2})\sin(\frac{\omega \Delta t}{2}) = (\frac{\omega_{pe}\Delta t}{2})^2.
\end{equation}
The roots of the dispersion relation are always real and therefore neither exponential growth nor damping is present at any choice of $\Delta t$. For this reason, the numerical scheme is linearly unconditional stable. For comparison, the numerical dispersion relation of explicit PIC method for a cold plasma leads to exponential growth for the well known condition $\omega_{pe}\Delta t > 2$~\cite{birdsall}. The dispersion analysis that includes grid effects is not carried out in the present paper, and it will part of a future work.

\begin{figure}
\centering
\includegraphics[width=\columnwidth]{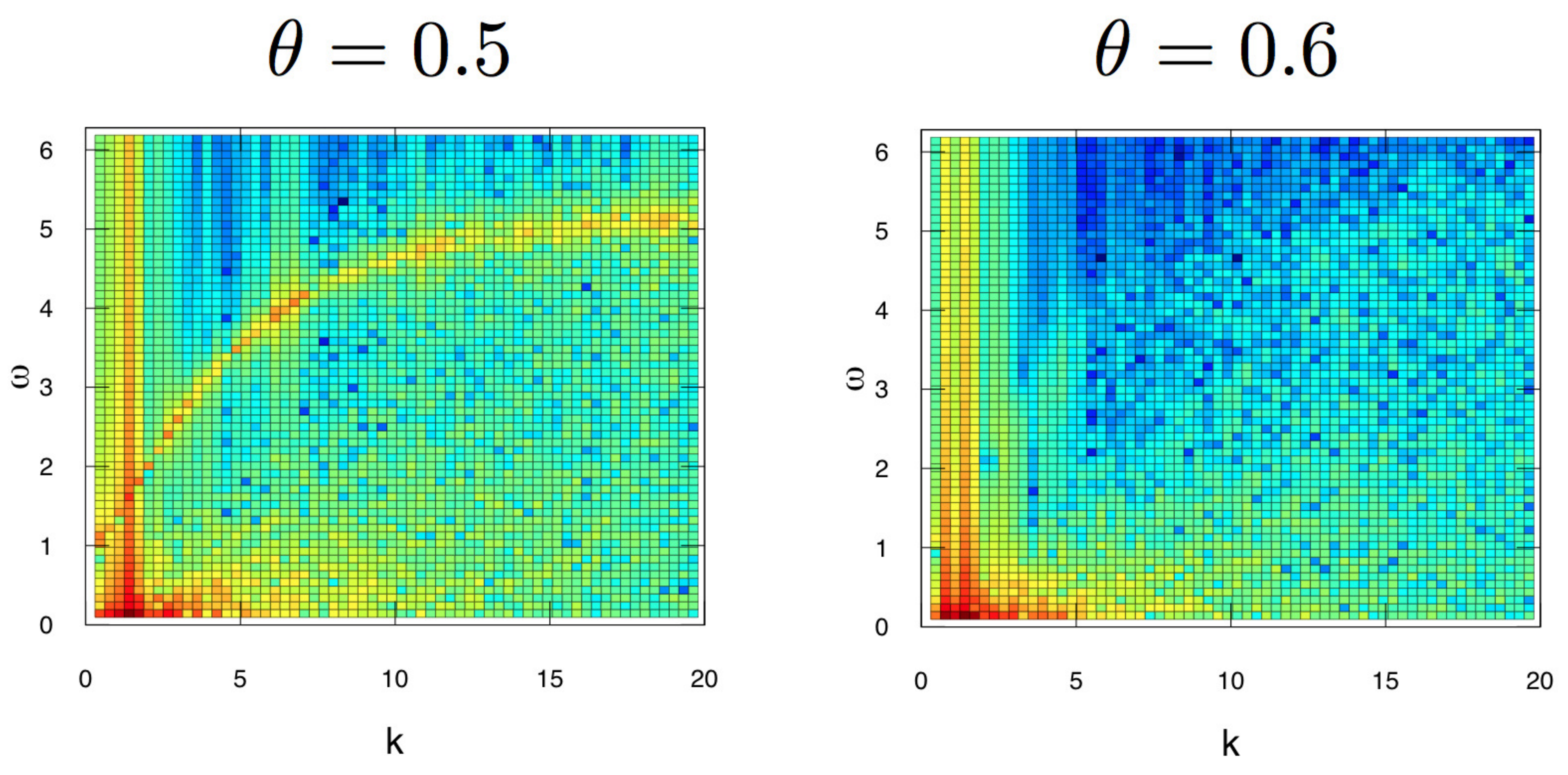}
\caption[]
{\label{Fig:DisperRelWeibelTheta}
Dispersion relation in the Weibel instability simulation for $\theta=0.5$ (energy conserving PIC code) and for $\theta=0.6$ (fully implicit PIC code with numerical damping). The Weibel instability is visible as vertical red lines in both simulations. The noise in the radiation field (a red line that is diagonal for small $k$ and becomes horizontal for high $k$) is present in the energy conserving simulation (left panel, $\theta=0.5$). Instead it is damped, and therefore not present, in the $\theta=0.6$ PIC simulation (right panel).}  
\end{figure}

In the case of implicit moment PIC methods, a $\theta$ parameter is introduced in the numerical scheme to decenter in time the discretization of the field equations~\cite{brackbill-cohen-85}. The quantities $q$ at time level $\theta$ are defined as $(1 - \theta)q^n + \theta q^{n+1}$. The numerical dispersion relation of the implicit moment PIC method has growing solutions for $\theta < 1/2$, damped solutions for  $\theta > 1/2$, and neither damping nor growth for $\theta = 1/2$~\cite{brackbill-cohen-85}. For $\theta>0.5$, the implicit moment PIC method damps high frequency waves, that are not resolved by the time step, as shown in Ref.\cite{brackbill-cohen-85}. On the contrary, in energy conserving PIC simulations unresolved waves are not artificially damped and hold over the simulation. To compare the behavior of other fully implicit PIC schemes with the energy conserving PIC method, a $\theta$ parameter has been introduced in the numerical scheme as follows: 
\begin{equation}
\begin{array}{l}
\tilde{\mathbf{v}}_p =\mathbf{v}_p^n+\frac{q_s\Delta t}{2m_s}\mathbf{{E}}^\theta_p\\
\mathbf{{v}}^\theta_p = \tilde{\mathbf{v}}_p+\frac{q_s\Delta
t}{2m_s c}\bigl(\tilde{\mathbf{v}}_p\times\mathbf{{B}}_p^\theta+\frac{q_s\Delta
t}{2m_s c}(\tilde{\mathbf{v}}_p\cdot\mathbf{{B}}_p^\theta)\mathbf{{B}}_p^\theta\bigr)/(1+\frac{q_s^2\Delta t^2}{4m_s^2c^2}{{B}^\theta_p}^2) \\
\mathbf{E}_g^{n+1} - \mathbf{E}_g^{n} = c \nabla \times \mathbf{{B}}^\theta_g \Delta t - 4 \pi \mathbf{{J}}^\theta_g \Delta t \\
\mathbf{B}_g^{n+1} - \mathbf{B}_g^{n} = - c \nabla \times \mathbf{{E}}^\theta_g \Delta t .
\end{array}
\end{equation}
The equations above are solved concurrently by a JFNK solver. For $\theta=0.5$, the method reduces to energy conserving PIC scheme, while for $\theta > 0.5$ does not conserve the total energy. In the latter case, the method artificially cools the plasma damping unresolved waves. This is clear from Figure \ref{Fig:DisperRelWeibelTheta}, where the numerical dispersion relation of a plasma undergoing Weibel instability~\cite{Weibel} is shown. The numerical dispersion relation has been calculated by applying the fast Fourier transform in space and in time to the $z$ component of the magnetic field, as shown in Ref.~\cite{Matsumoto}. The Weibel instability can be seen as a vertical red line in both panels of Figure \ref{Fig:DisperRelWeibelTheta}. In addition, thermal noise is visible as a red line, that is diagonal for small $k$ and becomes horizontal for high $k$, only in the left panel of Figure \ref{Fig:DisperRelWeibelTheta} for the energy conserving PIC code with $\theta=0.5$. The radiation field is due to aliasing errors, a feature of the quadratically conserving schemes~\cite{Arakawa:1990}. On the contrary, the radiation field noise is damped in the fully implicit PIC simulation with $\theta = 0.6$ and not visible in the right panel.

\section{Implementation}
An energy conserving PIC code has been developed in the Matlab/Octave programming language. For implementation simplicity, the code is 1D3V \cite{birdsall} with uniform grid: the space is one dimensional while the particle velocities, the electric and magnetic fields have three components. The extension of the code to the three dimensional case is straight-forward. 

After the simulation has been initialized setting self-consistently particle quantities and electromagnetic fields, and ensuring that the Gauss' law is satisfied initially, two steps are completed at each computational cycle, as shown in Figure \ref{fig:ECPICalgorithm}:
\begin{enumerate}
\item  The values of the dependent variables $\mathbf{E}_g^{n+1}$, $\mathbf{B}_g^{n+1}$ and $\mathbf{\bar{v}}_p$ are determined with the JFNK solver. Given $\mathbf{x}^n_p$, $\mathbf{v}_p^n$, $\mathbf{E}_g^{n}$ and $\mathbf{B}_g^{n}$ from the previous computational cycle, $\mathbf{\bar{v}}_p$, $\mathbf{E}_g^{n+1}$ and $\mathbf{B}_g^{n+1}$ are calculated solving concurrently Equations \ref{vn+1/2} and \ref{MaxwDiscrete} by a JFNK solver~\cite{Kelley:1995}.    
\item The new particle positions and velocities $\mathbf{x}_p^{n+1}$, $\mathbf{v}_p^{n+1}$ are updated with Equations \ref{dif_eom2}.
\end{enumerate}

The JFNK method solves the non-linear system $\mathbf{G}(\mathbf{x}) = \mathbf{0}$ iteratively by computing successive linear systems:
\begin{equation}
\label{NewtonIterations}
\left.{\frac{\partial \mathbf{G}( \mathbf{x})} {\partial\mathbf{x}}}\right |_i \delta \mathbf{x}_i = - \mathbf{G}(\mathbf{x}_i).
\end{equation}
the solution guess $\mathbf{x_i}$ at the iteration $i$ of the initial non-linear system $\mathbf{G}(\mathbf{x}) = \mathbf{0}$ is calculated as:
\begin{equation}
\label{update_x}
\mathbf{x}_{i+1} = \mathbf{x}_{i} + \delta \mathbf{x}_{i}.
\end{equation}
The solution of the linear system \ref{NewtonIterations} and the solution update (Eq. \ref{update_x}) compose the Newton iteration, that stops when: 
\begin{equation}
\parallel \mathbf{G}( \mathbf{x}_i) \parallel < \epsilon_a + \epsilon_r \parallel \mathbf{G}( \mathbf{x}_0) \parallel ,
\end{equation}
where $\parallel \cdot \parallel$ is the Euclidean norm, and $\epsilon_a$ and $\epsilon_r $ are the absolute and relative error tolerance values.
The successive linear systems \ref{NewtonIterations} are solved iteratively by a Krylov method, the Generalized Minimal Residual (GMRes) solver ~\cite{Kelley:1995} in the present study. The Krylov method convergence is adjusted at each Newton iteration as follow:
\begin{equation}
\parallel \left.{\frac{\partial \mathbf{G}( \mathbf{x})} {\partial\mathbf{x}}}\right |_i \delta \mathbf{x}_i  + \mathbf{G}(\mathbf{x}_i) \parallel < \zeta_i \parallel  \mathbf{G}(\mathbf{x}_i )\parallel ,
\end{equation}
where $\zeta_i$ is the inexact Newton parameter \cite{Kelley:1995}. The number of iterations of the GMRes solver are called Krylov iterations. The Jacobian ${\frac{\partial \mathbf{G}( \mathbf{x})} {\partial\mathbf{x}}}$ is not calculated directly, but instead the Gateaux derivative is used to compute:
\begin{equation}
\left.{\frac{\partial \mathbf{G}( \mathbf{x})} {\partial \mathbf{x}}}\right |_i \delta \mathbf{x}_i = \lim_{\epsilon\rightarrow 0} \frac{\mathbf{G}( \mathbf{x}_i + \epsilon \,\delta \mathbf{x}_i ) - \mathbf{G}( \mathbf{x}_i) }{\epsilon},
\end{equation}
where $\epsilon$ is a small but finite number. Because the the Jacobian does not need to be formed and calculated explicitly, the method is said "Jacobian-free".

A guess of the particle average velocity, of the new electric and magnetic fields (the dependent variables) is given initially as vector $\mathbf{x}_{KR}$ to the GMRes solver. A better estimate of these values is calculated by minimizing through successive solver iterations the residual $\mathbf{r}$ (the difference between the known term $\mathbf{b}$ and $A(\mathbf{x}_{KR})$, the non-linear system $A$ applied to the solution guess $\mathbf{x}_{KR}$)
\begin{equation}
\mathbf{r} = \mathbf{b} - A(\mathbf{x}_{KR}).
\end{equation}
A function where the residual $\mathbf{r}$ is calculated, must be provided to the JFNK solver. The residual is computed by solving the Equations \ref{vn+1/2} and \ref{MaxwDiscrete} for the problem unknowns, $\mathbf{\bar{v}}_p$ $\mathbf{E}_g^{n+1}$ and $\mathbf{B}_g^{n+1}$, in three successive steps:
\begin{enumerate}
\item Given a $\mathbf{\bar{v}}_p$ estimate in $\mathbf{x}_{KR}$, $\mathbf{\bar{x}}$ is calculated as $\mathbf{\bar{v}}_p\Delta t /2$, and $\mathbf{\bar{J}}_g$ is computed with Equation \ref{Jmedio}.
\item Given $\mathbf{E}_g^{n+1}$ and $\mathbf{B}_g^{n+1}$ estimates in $\mathbf{x}_{KR}$, $\mathbf{\bar{E}}_p$ and $\mathbf{\bar{B}}_p$ are calculated with Equation \ref{EBmediop}.
\item The residual $\mathbf{r}$ is computed with Equations \ref{vn+1/2} and \ref{MaxwDiscrete}.
\end{enumerate}
The solution of the particle equations of motion and field equations, and the current deposition are completed at each Krylov iteration.

\begin{figure}
\centering
\includegraphics[width=\columnwidth]{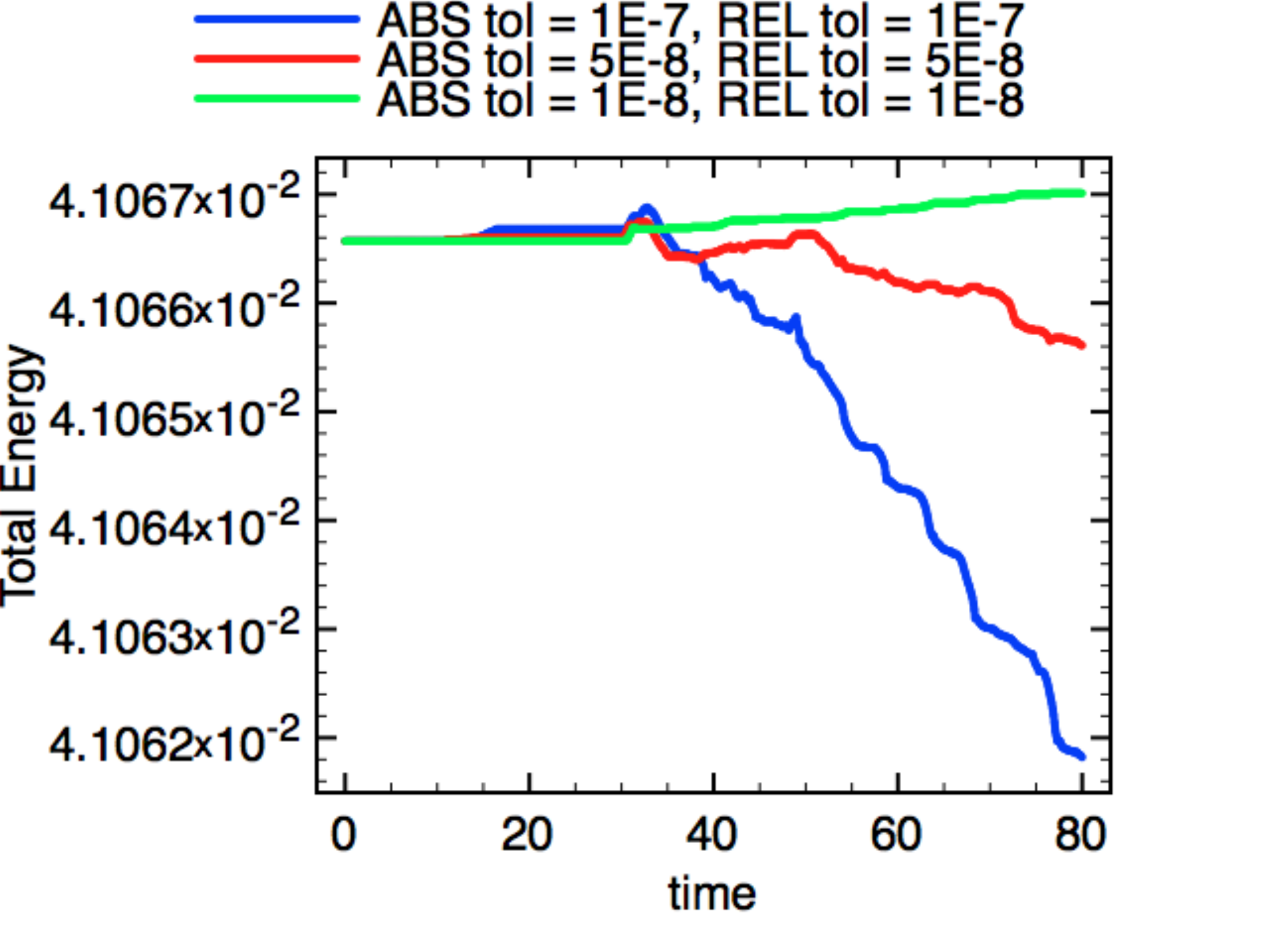}
\caption[]
{ \label{fig:TotalEnergyDifferentTol}
Comparison of energy histories in a two-stream instability simulation with different absolute and relative error tolerance values. Smaller tolerance errors lead to an increased energy conservation.}  
\end{figure}

A skeleton version of the Matlab/Octave code for the electrostatic limit with electrons and motionless ions is presented in Appendices A and B to show the simplicity of the proposed PIC method. The software implementation of the Newton Krylov GMRes solver is from the Kelley's textbook~\cite{Kelley:1995} and available at the website~\cite{MatlabCode}. In all the simulations, the solver maximum number of Newton and Krylov iterations is set to 40, and the inexact Newton parameter $\zeta_i$ is determined by the Eisenstat-Walker formula~\cite{Kelley:1995}. The Eisenstat-Walker parameter is chosen as 0.9. In the energy conserving PIC method, smaller error tolerance values lead to simulations with increased energy conservation. This is clearly visible in Figure \ref{fig:TotalEnergyDifferentTol}, where the energy history of the same simulation of the two-stream instability with different absolute and relative solver tolerance values is plotted. On the contrary, it has been found that decreasing the error tolerances does not have any effect in the conservation of the momentum. 

\section{Simulation Results}
The energy conserving PIC codes have been tested throughly. The goal of these tests is first to verify the new PIC method through comparison of the simulation results with analytical  theory, and second to show the exact energy conservation. In this paper, the energy conserving PIC code is first run in the electrostatic formulation for the problems of the finite grid and two-stream instabilities, and then in the fully electromagnetic case for the Weibel instability test.

\subsection{Finite Grid Instability}
An aliasing instability, called {\em finite grid instability}, arises in explicit momentum conserving PIC methods when the simulation grid spacing is approximately two times larger than Debye length $\Lambda_D$~\cite{birdsall,Matsumoto}. The finite-grid instability heats non physically the plasma, until the Debye length reaches a value comparable to half the grid spacing. Because this instability introduces numerical heat in the system, it appears as a macroscopic violation of the energy conservation. 

\begin{figure}
\centering
\includegraphics[width=\columnwidth]{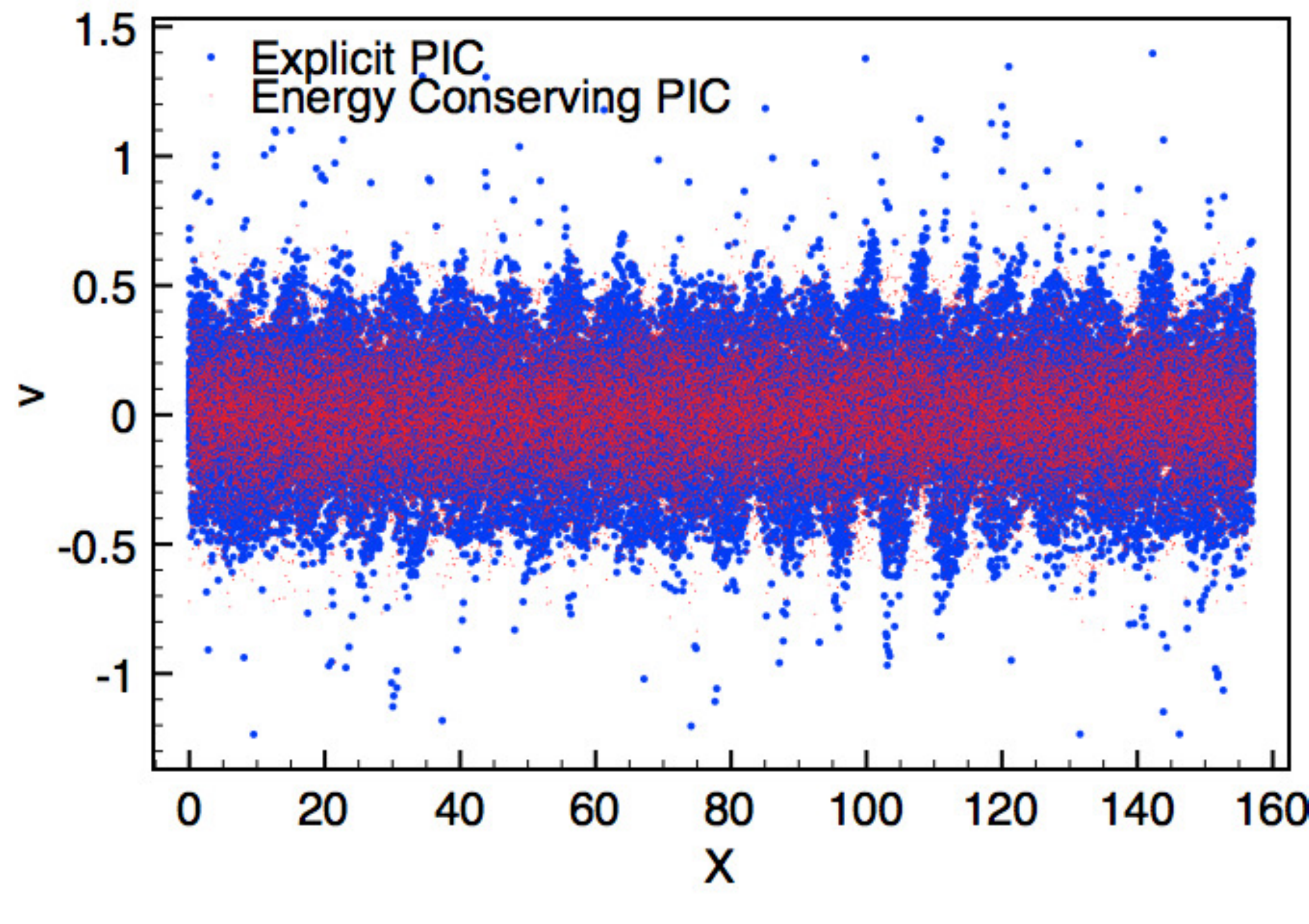}
\caption[]
{ \label{fig:FiniteGridPS}
Phase space plot of a Maxwellian plasma simulation at $t=100 \omega_{pe}^{-1}$ with the energy conserving (red dots) and explicit (blue dots) PIC codes. The finite grid instability appears as peaks of the electron distribution in the phase space in the explicit PIC simulation, while is not present in the energy conserving PIC simulation, where the plasma retains the initial Maxwellian distribution.}  
\end{figure}

\begin{figure}
\centering
\includegraphics[width=\columnwidth]{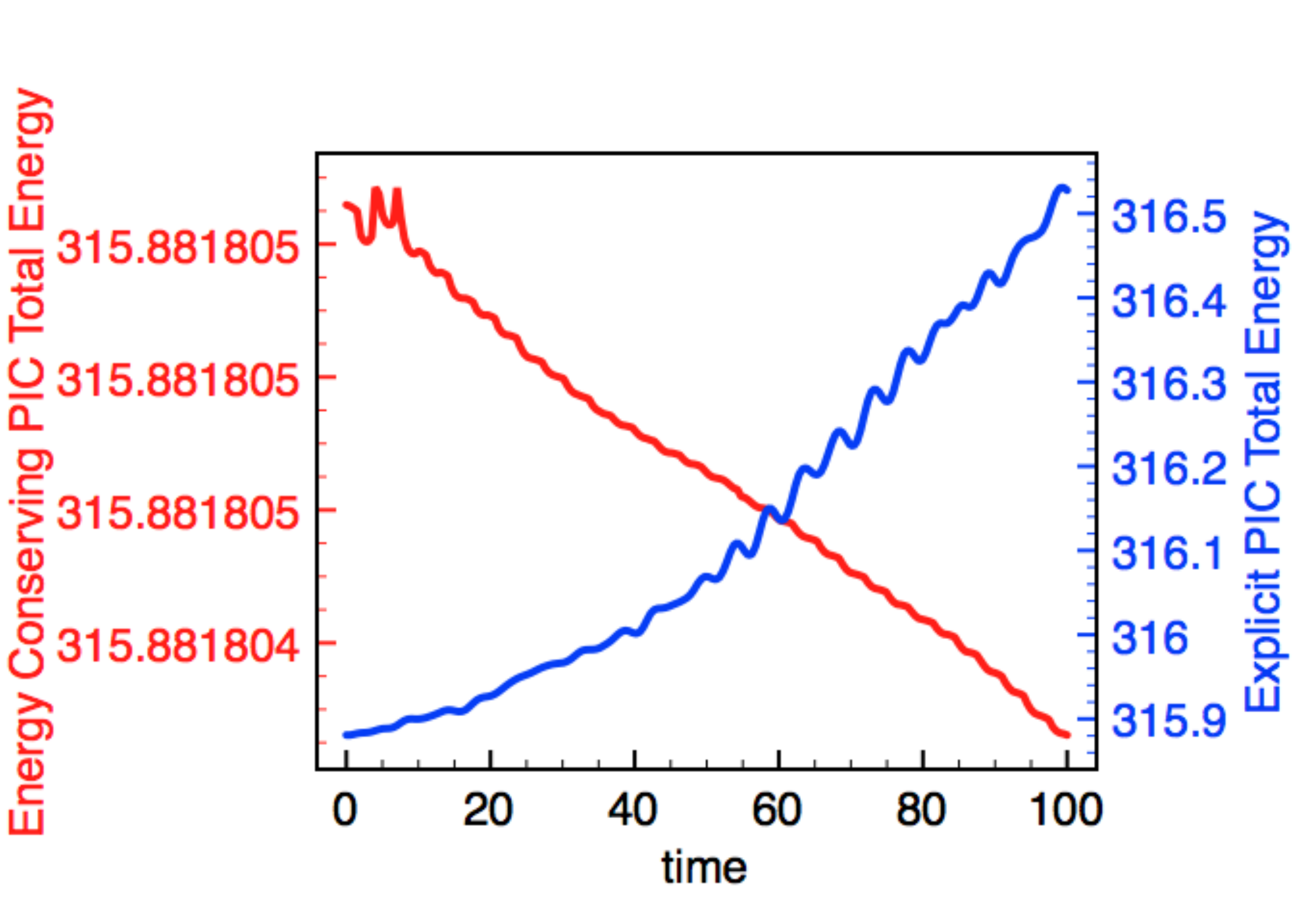}
\caption[]
{ \label{fig:FiniteGridEnergy}
Total energy history of a Maxwellian plasma simulation with the energy conserving PIC code (red line and left $y$ axis) and with the explicit momentum conserving PIC code (blue line and right $y$ axis). Energy is in $n_em_ec^2/2$ units. Because of the finite grid instability, the total energy in the explicit PIC simulation increases 2\%. On the contrary, the energy variation is limited to $10^{-7}$\% in the case of the energy conserving PIC simulation.}  
\end{figure}

To test the energy conserving electrostatic PIC method against the finite grid instability, a Maxwellian plasma is initialized with thermal velocity $v_{the}=0.2c$ in a simulation box $50\pi c/\omega_{pe}$ long with 64 grid cells and 50,000 particles. The Debye length $\Lambda_D = v_{the}/\omega_{pe} = 0.2 c/\omega_{pe}$ results approximately ten times smaller than the grid spacing $\Delta x =  2.45 c/\omega_{pe}$. This geometrical set-up leads to the finite grid instability if an explicit momentum conserving PIC code is used. The simulation evolves over 200 computational cycles with time step equal to $0.5 \omega_{pe}^{-1}$. Therefore, the numerical constraint $v_{the}\Delta t/\Delta x = 0.245 < 1.5 $ is satisfied and numerical instability does not arise because of the non conservation of momentum. The absolute and relative solver error tolerance values are both set to $10^{-8}$.  In addition, a simulation with an explicit momentum conserving PIC code, starting from the same initial configuration, has been run to compare the results. Figure \ref{fig:FiniteGridPS} represents the phase space (each dot represents a particle in the position-velocity space) of the system at $t=100 \omega_{pe}^{-1}$ for the energy conserving (red dots) and explicit momentum conserving (blue dots) methods. The finite grid instability in the simulation with the explicit PIC code is visible from the blue peaks in the phase space. Instead the plasma retains the initial Maxwellian distribution in the energy conserving PIC simulation. Figure \ref{fig:FiniteGridEnergy} shows the total energy history for the two simulations. The finite grid instability produces a 2\% energy increase in the explicit PIC simulation, while the variation with the energy conserving PIC method is very low, $10^{-7}$\%.
Because the energy conserving PIC does not undergo finite grid instability, and does not require to resolve the Debye length, the proposed PIC method is well suited for simulations with large domains and/or few grid points. A complete linear and non-linear analysis of the finite grid instability in the energy conserving PIC has not been carried out and it will be a topic of a future work.

\subsection{Two-stream Instability}
The two-stream instability is an important phenomenon occurring in space physics, in the injection systems for nuclear fusion machine, and in particle accelerators~\cite{Krall}. In this problem, two electron beams move initially in opposite directions. The two beams extinguish as result of the beams instability. A simulation of the two-stream instability has been completed with the energy conserving PIC code. The drift velocity of the counter-streaming electron cold beams is $\pm0.2\,c$; the simulation box is 2.053 $c/\omega_{pe}$ long with 64 grid points and periodic boundaries. The number of electrons and ions is 200,000. The charge to mass ratio for electrons and ions is one and 1836 respectively. The simulation time step $\Delta t$ is 0.1 $\omega_{pe}^{-1}$. This set-up leads to $ v_c \Delta t /\Delta x = 0.623 < 1.5$, and thus the instability due to the non conservation of the total momentum does not occur. The absolute and relative tolerance are both set to $10^{-8}$. 

\begin{figure}
\centering
\includegraphics[width=\columnwidth]{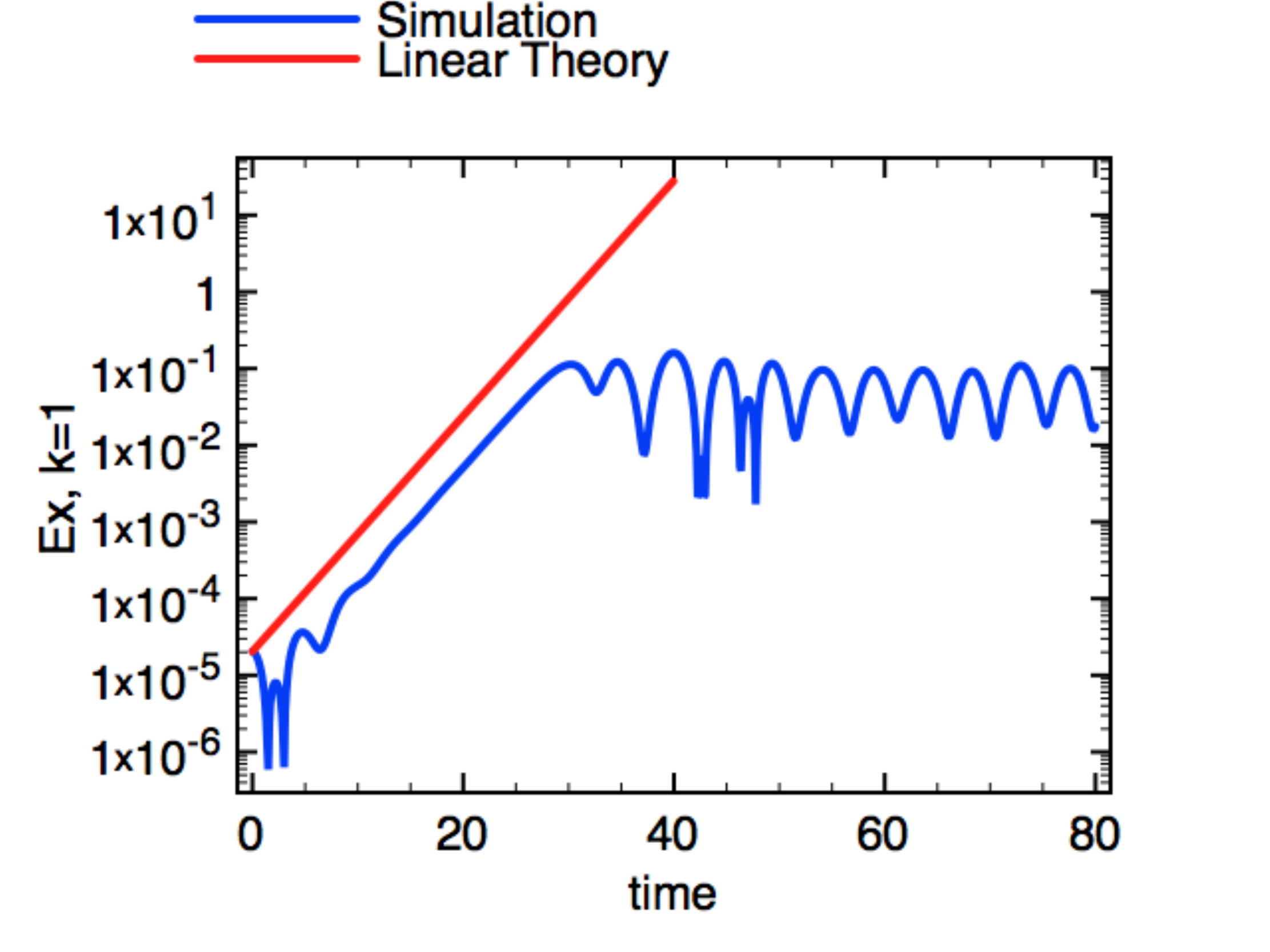}
\caption[]
{ \label{fig:2streamCompLin}
Comparison between linear theory and energy conserving PIC simulation of the two-stream instability. The $k=1\,\omega_{pe}/c$ spectral component of the electric field (in blue color) grows as predicted by the linear theory (red line). The electric field is in $\sqrt{4 \pi n_e m_e c^2}$ units.}  
\end{figure}

\begin{figure}. 
\centering
\includegraphics[width=\columnwidth]{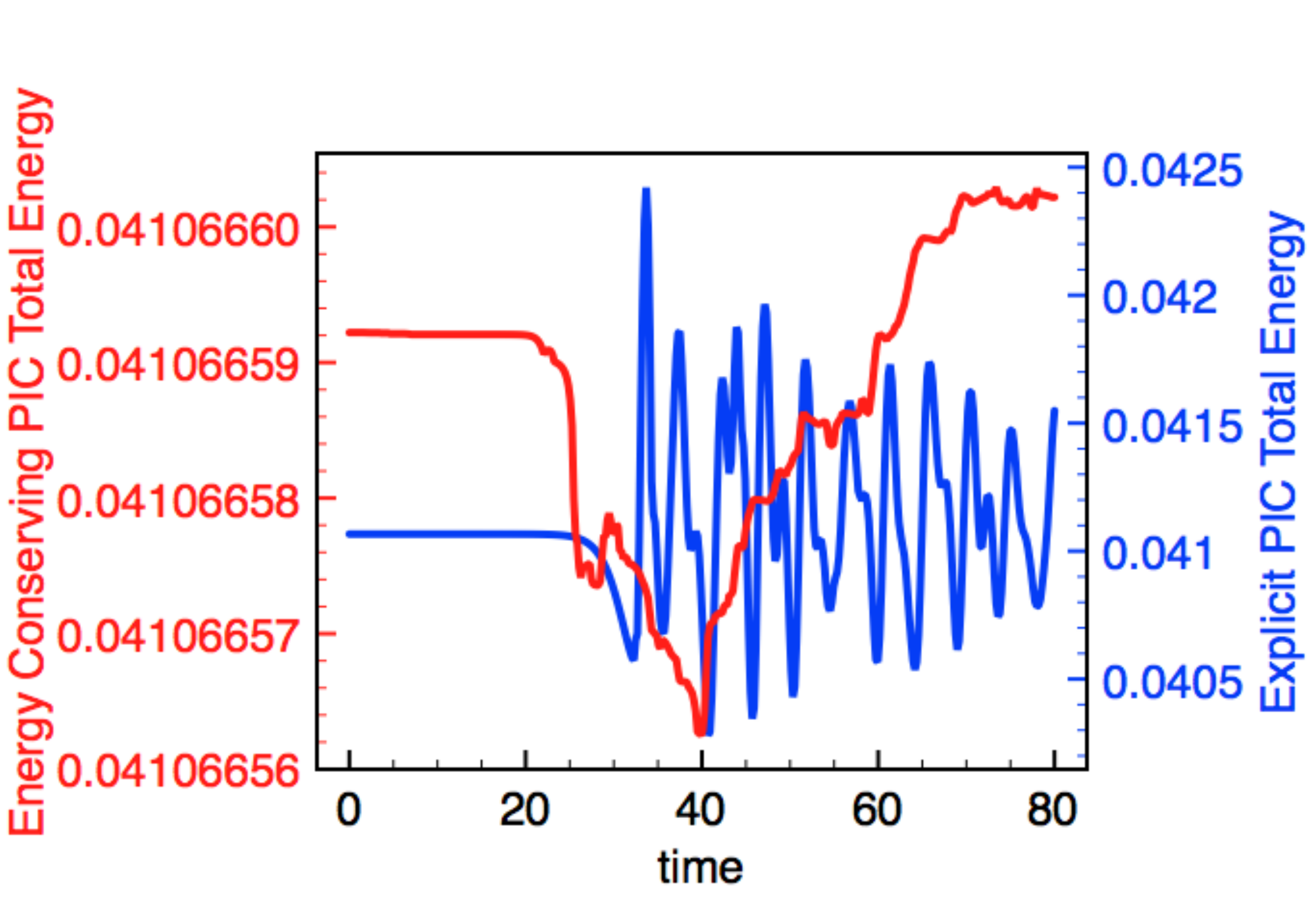}
\caption[]
{ \label{fig:CompExpEC}
Comparison of the total energy history of the explicit momentum conserving (blue line, right $y$ axis) and energy conserving PIC (red line, left $y$ axis) codes for the two-stream instability. Energy is in $n_em_ec^2/2$ units. The plot shows 5\% and $10^{-4}\;\%$ variations for the explicit and energy conserving PIC codes respectively.}  
\end{figure}

The linear theory predicts a growth rate of instability for the spectral component $k=1\,\omega_{pe}/c$ equal to 0.35355 $\omega_{pe}$~\cite{Krall} in this system configuration. Figure \ref{fig:2streamCompLin} shows an excellent agreement between the linear theory in red line and the simulation results in blue line in the linear stage of the instability.
The energy variation for an explicit momentum conserving, and an energy conserving PIC code, simulating the two-stream instability are compared in Figure \ref{fig:CompExpEC}. The total energy in the explicit PIC code shows an approximately 5\% variation, while  the variation is limited to $10^{-4}\;\%$ in the energy conserving PIC code. An increased energy conservation can be achieved, decreasing the error tolerance values, as shown in Figure \ref{fig:TotalEnergyDifferentTol}.

\subsection{Weibel Instability}
The Weibel instability is triggered by an anisotropic temperature in the plasma~\cite{Weibel,Morse:1971}. The instability converts the plasma temperature anisotropy into magnetic field relaxing the initial particle distribution function to an isotropic Maxwellian. Because plasma temperature anisotropies are very common in laboratory, space and astrophysical plasmas, the Weibel instability has been thoroughly studied with PIC methods~\cite{Morse:1971}. The energy conserving PIC code has been tested in a simulation box $2 \pi c/\omega_{pe}$ long, with 64 grid points and periodic boundaries. The time step $\Delta t$ is 0.25 $\omega_{pe}^{-1}$ and simulation is run for 400 computational cycles. 100,000 electrons are initialized with uniform distribution in space and bi-Maxwellian distribution with thermal velocity $v_{thy} = 0.4$, and anisotropy $a = (v_{thy}^2/v_{thx}^2-1) = 15$. Thus, the condition $v_{thx}\Delta t / \Delta x = 1.02$ to avoid the aliasing instability due to the non conservation of momentum, is satisfied. The mass ratio between ions and electron is 1836, and ions are initialized with same temperature of electrons. The electric and magnetic field are initially zero. The solver absolute and relative error tolerance values are both set to $10^{-5}$ .

\begin{figure}
\centering
\includegraphics[width=\columnwidth]{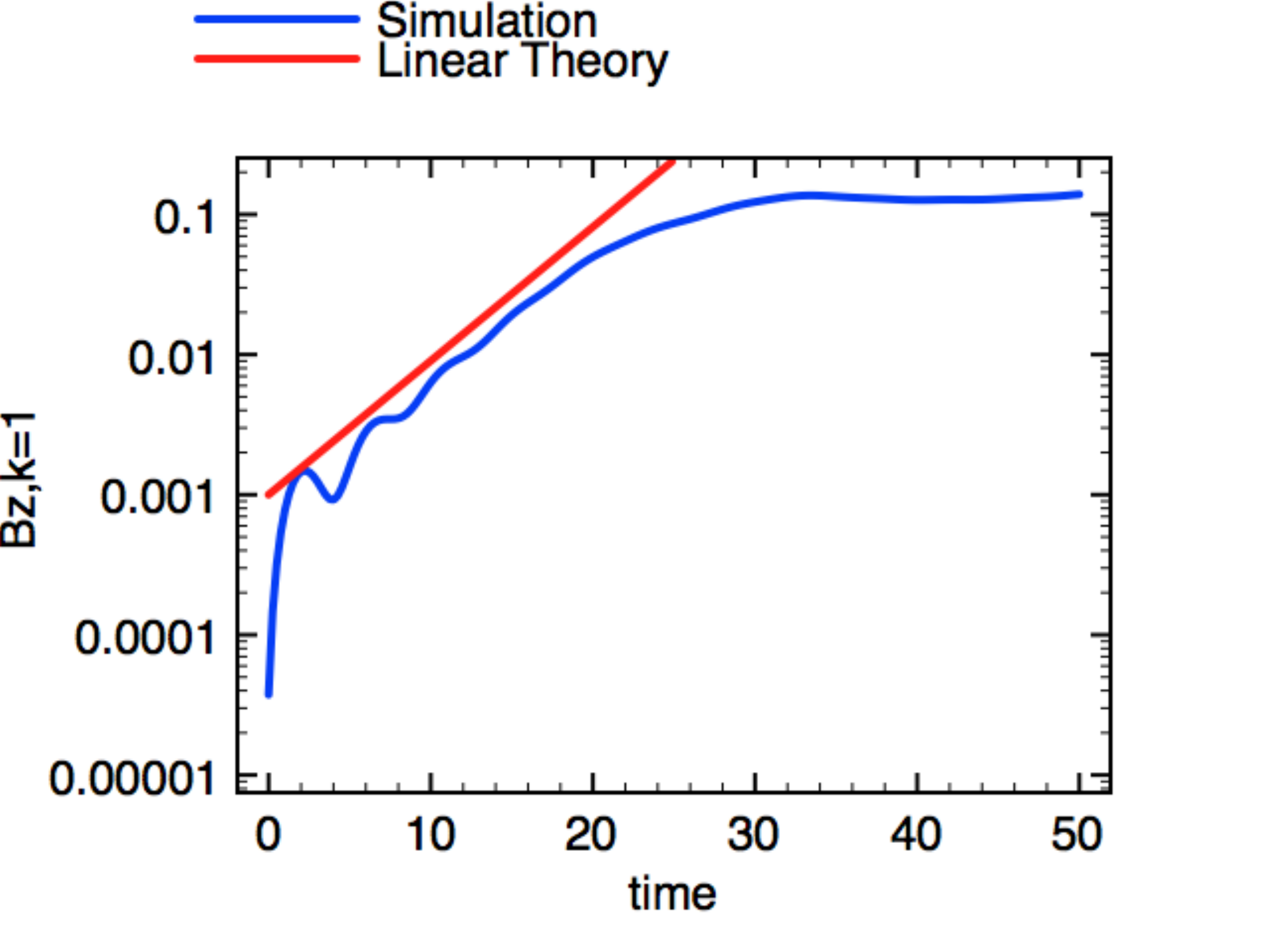}
\caption[]
{ \label{fig:WeibelLinTheory}
Comparison between the spectral component $k=1\omega_{pe}/c$ of $B_z$ and linear theory. Simulation and analytical results are in good agreement in the linear regime of the Weibel instability. The magnetic field is in $\sqrt{4 \pi n_e m_e c^2}$ units.}  
\end{figure}

\begin{figure}
\centering
\includegraphics[width=\columnwidth]{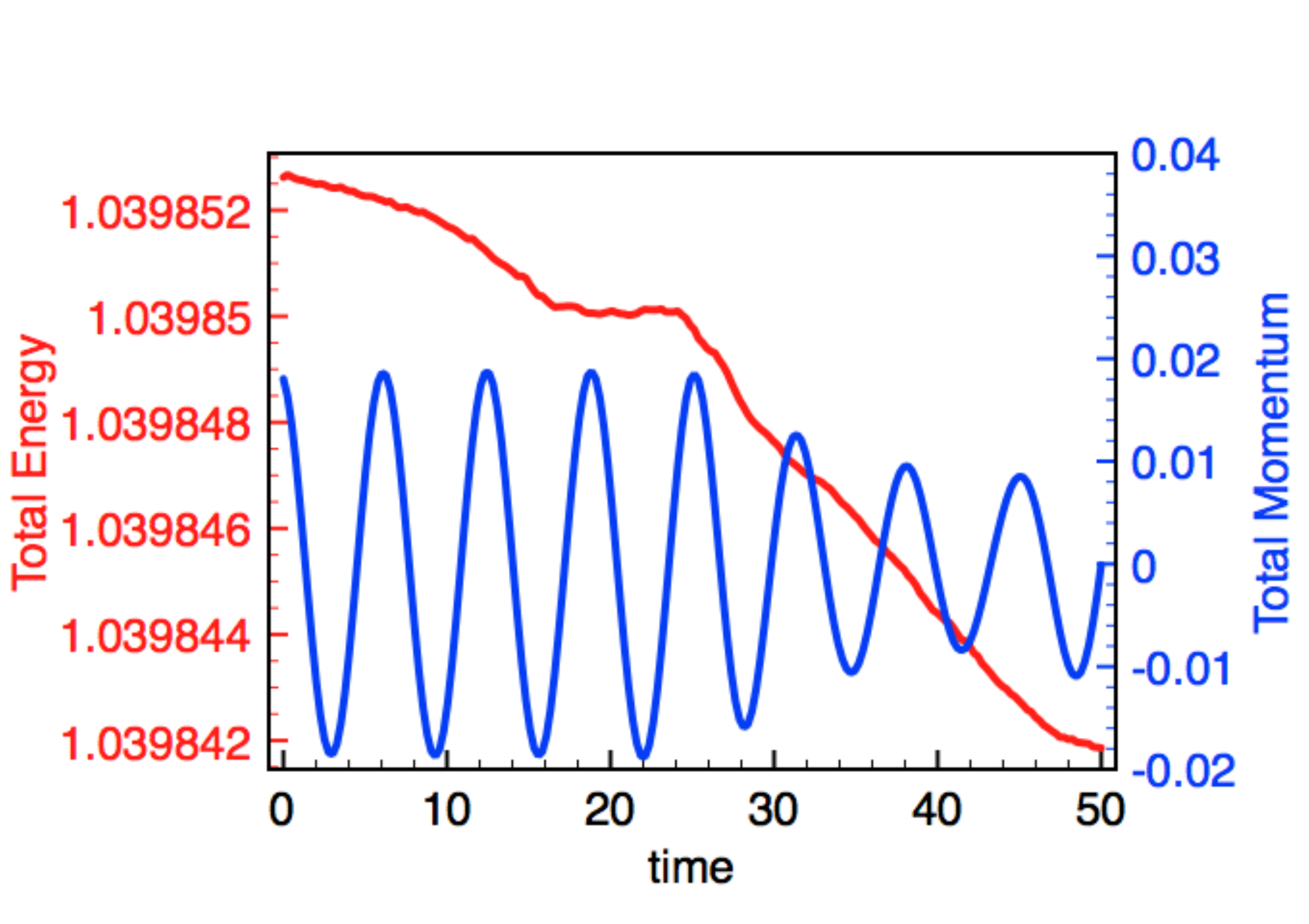}
\caption[]
{ \label{fig:WeibelEnergyMomentum}
Total energy and momentum histories in the Weibel instability simulation. Energy and momentum are in $n_em_ec^2/2$ and $m_ec$ units. The left red $y$ and the right blue $y$ axes correspond to the total energy and momentum respectively. The total energy is conserved within $10^{-3}\;\%$ variation, while the momentum is not conserved and oscillates between $-0.02 m_ec$ and $0.02 m_ec$.}  
\end{figure}

The linear theory predicts a growth rate of the $B_z$ component for the spectral component $k=1\omega_{pe}/c$ equal to $0.22 \omega_{pe}$~\cite{Krall}. Figure \ref{fig:WeibelLinTheory} compares the Weibel instability simulation results with the analytical calculations. Linear theory and simulation results are in good agreement in the linear regime of the instability. However, the growth of the magnetic field is not purely exponential as predicted~\cite{Krall}, but presents an oscillation in time. This oscillation is caused by the radiation field noise~\cite{Morse:1971,Langdon:1972}. As shown in Section 5 and in the left panel of Figure \ref{Fig:DisperRelWeibelTheta}, the numerical dispersion relation of the Weibel instability with the energy conserving PIC code shows the presence of waves due to the thermal noise \cite{Langdon:1972}.
Figure \ref{fig:WeibelEnergyMomentum} shows the total energy and momentum history in a Weibel instability simulation with the energy conserving PIC method. The total energy variation is within $10^{-3}\;\%$. Momentum is not conserved and oscillates between $-0.02m_ec$ and $0.02 m_ec$.

\section{Performance Results}
A study of the computational performance of the proposed PIC scheme has been completed. A Maxwellian plasma, composed of electron and a background ions, is simulated with an electrostatic energy conserving PIC in  a $6.4 c/\omega_{pe}$ long box over a period of 1000 cycles. The other simulation settings vary to perform a parametric study. Table \ref{my_table} shows the average of Newton and Krylov iterations, and the execution time for different number of grid points, time step, number of particles, absolute and relative error tolerance values. The tests have been completed on a 2.4 GHz Intel Core Duo, 2 GB RAM memory, Mac OS X 10.6.6  using the Matlab 7.5 and the code in Appendices A and B.
\begin{table}
\caption{Average number of Newton and Krylov (GMRes) iterations and execution time for different number grid points, time step, number of particles, and solver error tolerances ($\epsilon_a$, $\epsilon_r$) for a simulation of Maxwellian plasma with an electrostatic energy conserving PIC code.}
\begin{center}
\begin{tabular}{c  c  c  c  c c c}

  \hline
  $N_g$ & $dt$ ($\omega_{pe}^{-1}$) & $N_s$  & $\epsilon_a$, $\epsilon_r$ &Newton & Krylov & Time (s)\\
   \hline
  64 & 0.1 & 100000 &  $10^{-7}$, $10^{-7}$&3.62 & 2.32 & 341.86\\
  {\bf 128} & 0.1 &  100000 &$10^{-7}$, $10^{-7}$& 4.04 & 2.33 & 401.28\\
  {\bf 256} & 0.1 & 100000 &$10^{-7}$, $10^{-7}$& 3.98 & 2.50 & 416.66\\
  {\bf 512} & 0.1 &  100000 &$10^{-7}$, $10^{-7}$& 4.02 & 2.49 & 413.18\\
  64 & {\bf 0.2} & 100000 &$10^{-7}$, $10^{-7}$& 4.15 & 2.45 & 408.38\\
  64 & {\bf 0.4} & 100000 &$10^{-7}$, $10^{-7}$& 4.03 & 2.58 & 424.98\\
  64 & {\bf 0.8} & 100000 &$10^{-7}$, $10^{-7}$& 4.31 & 2.69 &455.57\\
  64 & 0.1 & {\bf 200000} &$10^{-7}$, $10^{-7}$& 3.25 & 2.37 & 617.13\\
  64 & 0.1 & {\bf 400000} &$10^{-7}$, $10^{-7}$& 2.91 & 2.39 & 1,144.90\\
  64 & 0.1 & {\bf 800000} &$10^{-7}$, $10^{-7}$& 2.79 & 2.42 & 2,194.00\\
  64 & 0.1 & 100000 & $\mathbf{10^{-8}}$, $\mathbf{10^{-8}}$&4.55 & 2.34 & 402.19\\
  64 & 0.1 & 100000 & $\mathbf{10^{-9}}$, $\mathbf{10^{-9}}$&4.82 & 2.42 & 433.95\\
  64 & 0.1 & 100000 & $\mathbf{10^{-10}}$, $\mathbf{10^{-10}}$&4.92 & 2.43 & 447.01\\
   \hline
\end{tabular}
\end{center}
\label{my_table}
\end{table}
The computational performance of the energy conserving weakly depends on the number of cells: increasing the number of grid point from 64 to 512 leads to a 20\% computational time increase. The time step has a similar effect on performance also. The simulation with $dt=0.8\omega_{pe}^{-1}$ takes only an additional 33\% computational time of the simulation with $dt=0.1\omega_{pe}^{-1}$. Instead, the computational performance strongly depends on the number of particles. In fact, doubling the number of particles leads to doubling the computational time. In addition, increasing the number of particles reduces the statistical noise, and consequently the convergence in the Newton step. The decrease JFNK error tolerance increases the degree of conservation energy as shown in Figure \ref{fig:TotalEnergyDifferentTol} at the cost of an increased number of Newton iterations and computational time.

\subsection{Kinetic Enslavement}
\begin{figure}
\centering
\includegraphics[width=\columnwidth]{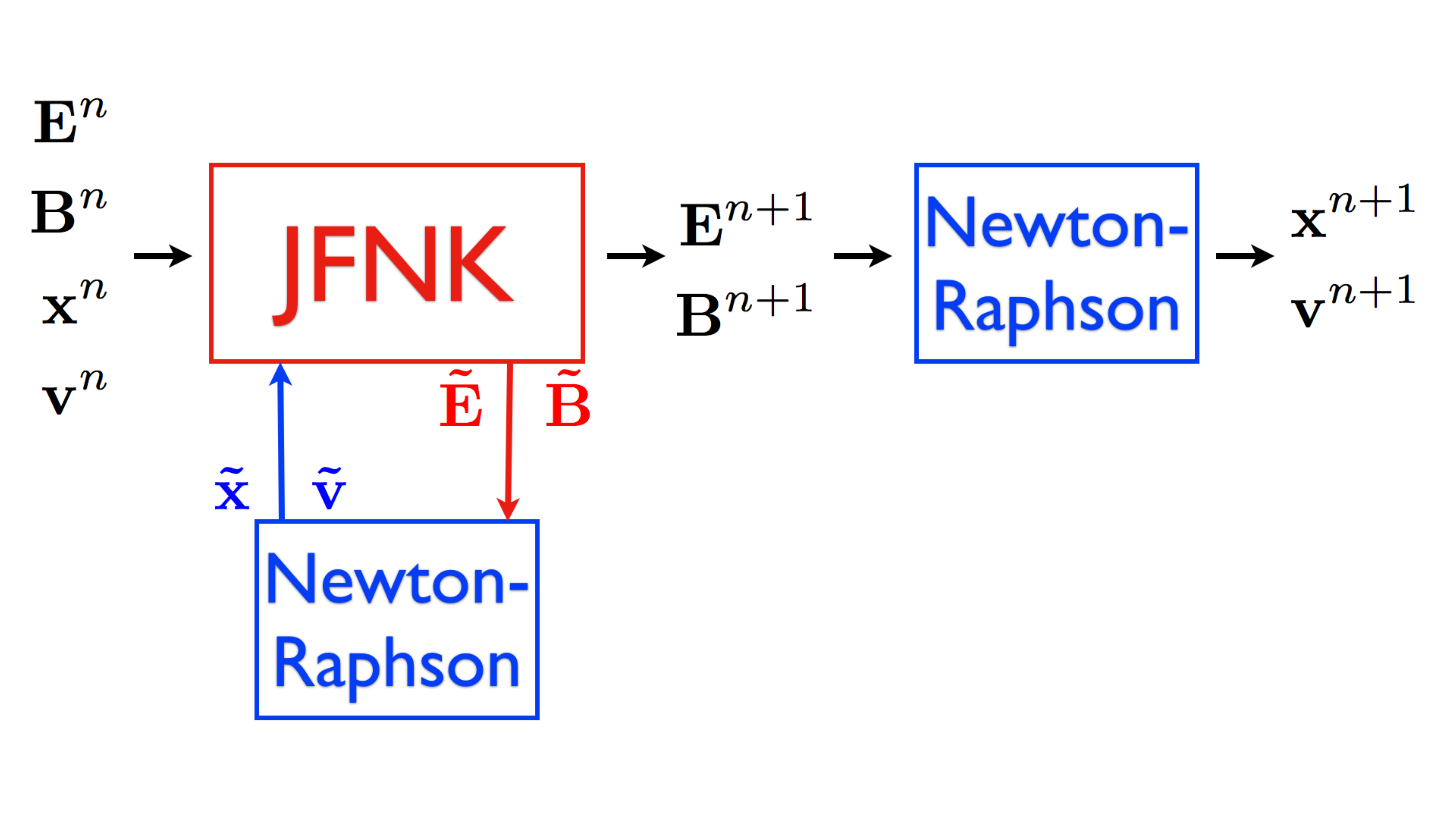}
\caption[]
{\label{Fig:JFNK_2}
Computation cycle for the energy conserving PIC with kinetic enslavement. Particle positions and velocities, $\tilde{\mathbf{x}}$, $\tilde{\mathbf{v}}$,  are calculated consistently with the fields, $\tilde{\mathbf{E}}$, $\tilde{\mathbf{B}}$ by Newton-Raphson method at each JFNK iteration.}  
\end{figure}

The main disadvantage of the energy conserving PIC method is that it requires the solution of a very large system whose size increases with the number of particles. The number of particles is easily more million in a typical PIC simulation, leading to matrix to be inverted whose rank size is of the order of million. To reduce the size of this matrix, it is possible to use a technique, called {\em kinetic enslavement}, following Ref.\cite{taitanoMSthesis}. In this method, the unknowns of the problem are only the value of the electric magnetic field on the grid points at the new time level and the JFNK solver computes only the field Equations \ref{MaxwDiscrete}. The particle equations of motion are calculated by a Newton-Raphson method and embedded in the field solver as function evaluations \cite{taitanoMSthesis}. Figure \ref{Fig:JFNK_2} shows the the computational cycle of the energy conserving PIC method with kinetic enslavement.  A guess of the electromagnetic field ($\tilde{\mathbf{E}}$, $\tilde{\mathbf{B}}$ in Figure \ref{Fig:JFNK_2}) is provided at each Newton iteration. Particle positions and velocities ($\tilde{\mathbf{x}}$, $\tilde{\mathbf{v}}$ in Figure \ref{Fig:JFNK_2}) are computed consistently with the the electromagnetic field guess by the Newton-Raphson method. 

\begin{figure}
\centering
\includegraphics[width=\columnwidth]{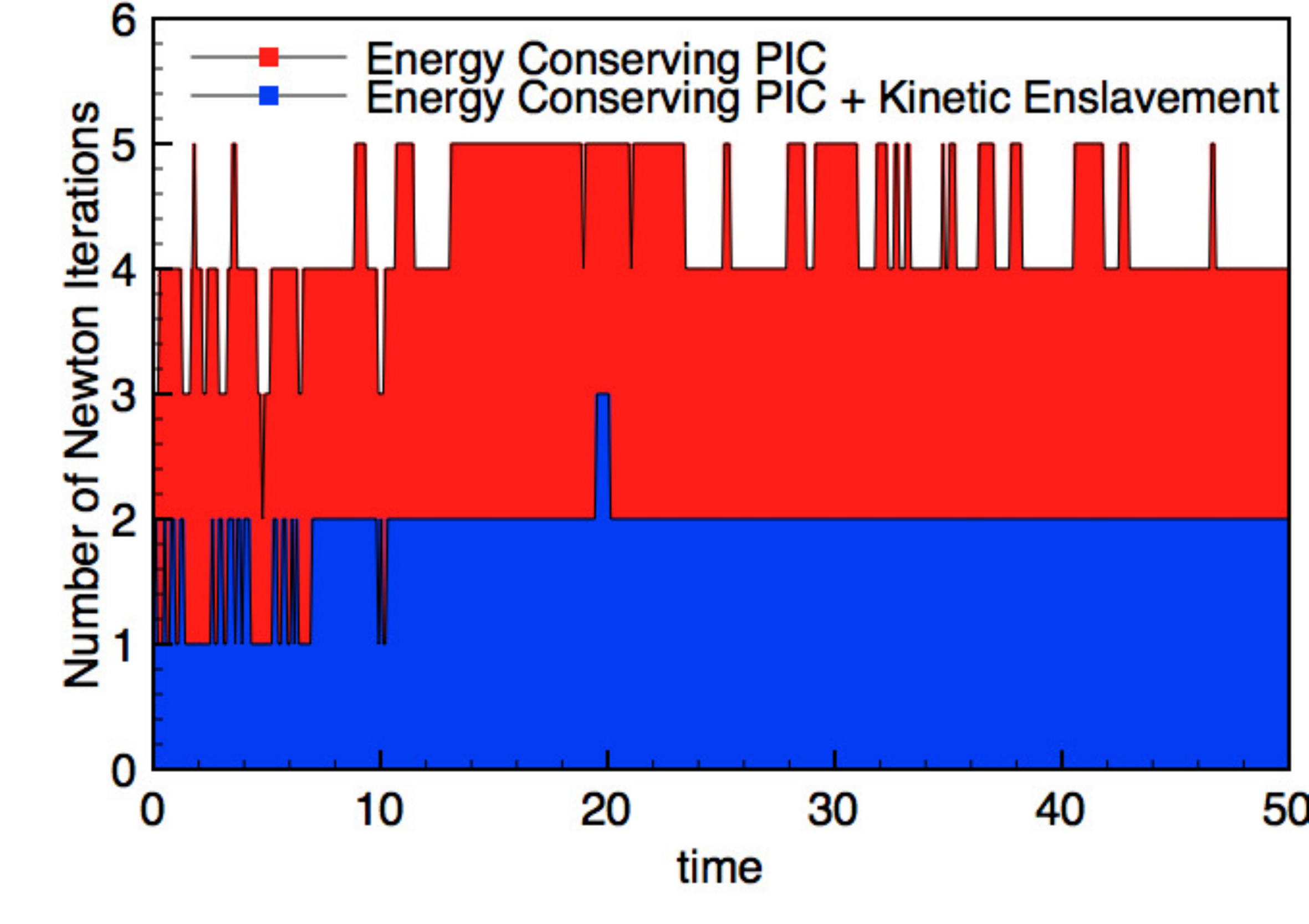}
\caption[]
{ \label{fig:IterationsNewton}
Comparison of number of Newton iterations in the energy conserving PIC code with and without kinetic enslavement for the two stream instability test.}  
\end{figure}

\begin{figure}
\centering
\includegraphics[width=\columnwidth]{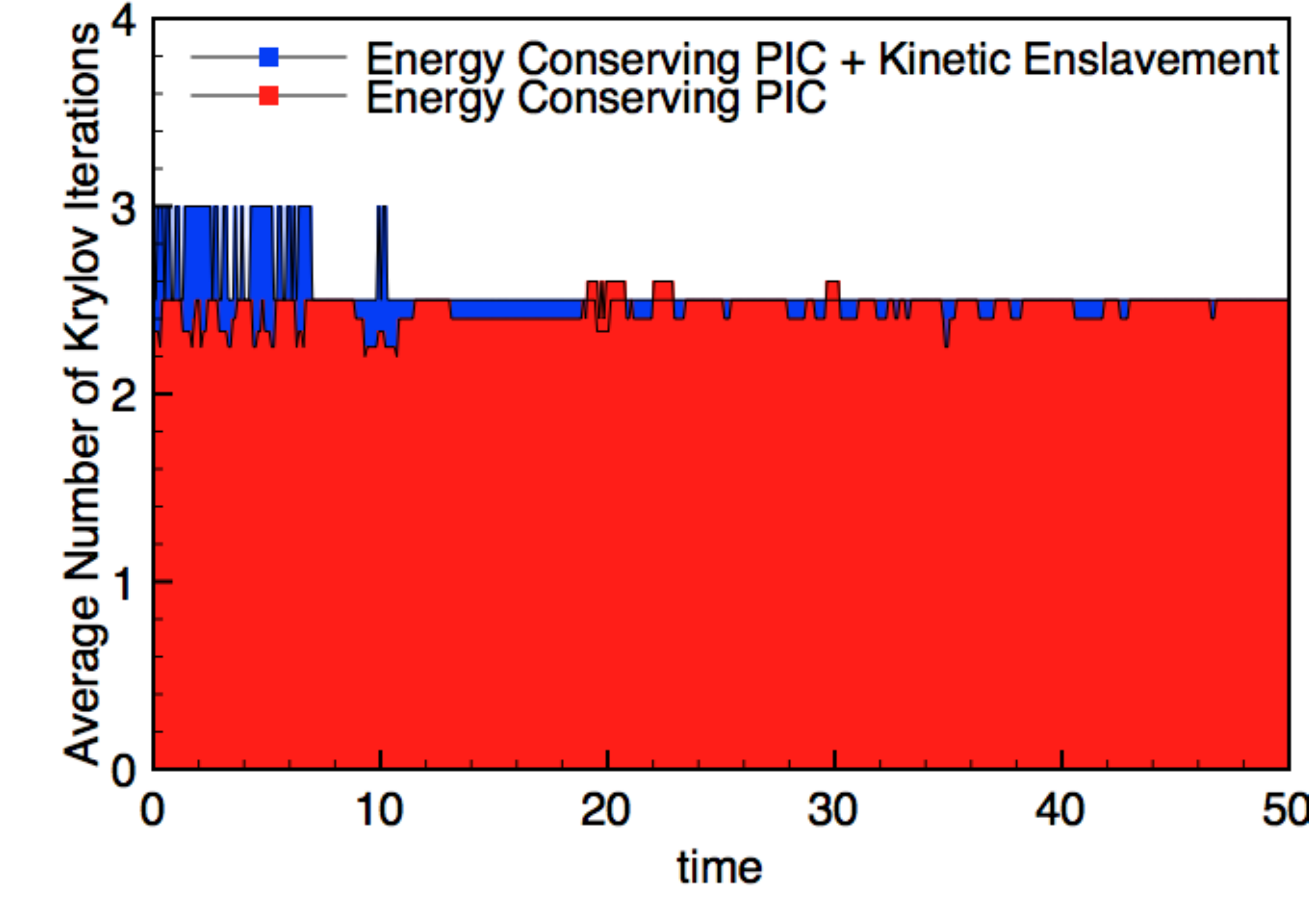}
\caption[]
{ \label{fig:IterationsKrylov}
Comparison of the average number of Krylov iterations per Newton step in the energy conserving PIC code with and without kinetic enslavement for the two stream instability test.}  
\end{figure}

An energy conserving PIC code with kinetic enslavement technique has been developed to test its effectiveness. Figures \ref{fig:IterationsNewton}, \ref{fig:IterationsKrylov} show a comparisons of the solver iterations for the energy conserving PIC code with and without the kinetic enslavement. The plots represent the number of Newton and average Krylov iterations per Newton step in the simulation of the two stream instability with an electrostatic energy conserving PIC code with and without kinetic enslavement. The two stream instability is simulated for 500 cycles, with $dt=0.1$, 64 grid points and 1000 particles. The Newton-Raphson convergence tolerance is set to $10^{-4}$, while the JFNK error tolerance values are set to $10^{-7}$. The size of the systems is reduced from $1064 \times 1064$ (energy conserving PIC) to $64 \times 64$ (energy conserving PIC with kinetic enslavement). In addition, it is clear from Figure \ref{fig:IterationsNewton} that the use of kinetic enslavement technique reduces the number of Newton iterations: the average number of Newton iterations is 1.91 and 4.33 in the simulation with and without kinetic enslavement method. The number of Krylov iterations remains almost unchanged in the two cases (Figure \ref{fig:IterationsKrylov}). It must be noted that an average of 3.95 Newton-Raphson iterations have been completed at each solver iteration. 

\section{Conclusions}
The algorithm, the properties, the implementation, the simulation and performance results of the energy conserving Particle-in-Cell method have been presented. The proposed PIC method has been tested against the finite grid, two-stream and Weibel instabilities to prove the algorithm correctness and its property of conserving exactly the total energy. The method is based on a non conservative definition of the current density. The numerical error due to the violation of the Gauss' law built up slowly in all the completed simulations and did not affect the results. The new method does not conserve the momentum, and a condition on the maximum number of cells a particle can cross per time step, must be satisfied to avoid an aliasing instability. The new PIC scheme is based on the implicit discretization of the governing equation, and therefore results linearly unconditionally stable. 

The energy conserving PIC method is a fully implicit PIC method\cite{Kim2005,markidisPHDthesis}, where the particle average velocities and the electromagnetic field are calculated self-consistently in the JFNK solver to preserve the system total energy. The performance of the fully implicit PIC methods, and a comparison between fully implicit and implicit moment PIC methods in terms of efficiency and required computational resources, have been presented in Refs.\cite{markidisPHDthesis,Kim2005}. It has been shown in this paper that the energy conserving PIC performance critically depends on two factors: the number of computational particles and the solver error tolerance values. In fact, the computational time increases linearly with the number of particles, and it is rather insensitive to the number of grid points and the time step. In addition, a smaller error tolerance value leads to a larger number of iterations, and therefore to a larger computational time. The use of the kinetic enslavement technique \cite{taitanoMSthesis} reduces the size of problem matrix and has the beneficial effect of decreasing the number of Newton iterations. 

\section*{Acknowledgment}
The authors are grateful to Gianni Coppa for his clever implementation of explicit electrostatic PIC in Matlab/Octave as well as for the fruitful discussions on the mathematical foundations of the PIC method. The authors are also grateful to Jerry Brackbill for the stimulating exchanges of ideas on the implicit PIC method. The present work is supported by the Onderzoekfonds KU Leuven (Research Fund KU Leuven) and by the European CommissionÕs Seventh Framework Programme (FP7/2007-2013) under the grant agreement no. 218816 (\url{soteria-space.eu}).

\providecommand{\noopsort}[1]{}\providecommand{\singleletter}[1]{#1}%


\appendix
\section{ECpicES.m}
A skeleton version of the energy conserving PIC in Matlab/Octave programming language is here presented for the electrostatic limit (Section 2.1) with a plasma of electrons and motionless background ions~\cite{MatlabCodeMarkidis}. The energy conserving PIC code requires additional files ({\em nsolgm.m}, {\em fdgmres.m}, {\em givapp.m}, and {\em dirder.m}), that are available at the Kelley's textbook website~\cite{MatlabCode}. 

After the initial parameters are set and the electric field is calculated solving the Gauss' law to ensure the charge conservation, the average particle velocities and the new electric field are calculated at line 52, and the new particle positions and velocities are updated at lines 55 and 56 at each computational cycle.
\begin{lstlisting}
% Energy Conserving PIC code 
global L; global dx; global NG; global DT; global N; global WP; 
global QM; global Q; global rho_back; 
global x0; global v0; global E0;

% simulation parameters
L=2*pi/3.0600; % Simulation box length
DT=0.1; % time step 
NT=800; % number of computational cycles
NG=128; % number of cells
N=500000; % number of particles
WP=1; % plasma frequency
QM=-1; % electron charge to mass ratio
V0=0.2; % beam velocity
VT=0.0; % thermal velocity
tol = [1E-7, 1E-7]; % absolute and relative error tolerance 
dx=L/NG; % grid spacing
Q=WP^2/(QM*N/L); % computational particle charge
rho_back=-Q*N/L; % background ion charge density 
histEnergy = [];  %total energy history
% two-stream instability
% initial particle positions
x0=linspace(0,L-L/N,N)'; % uniform in space
% initial particle velocities
v0=VT*randn(N,1); % two counterstreaming beams
pm=[1:N]'; pm=1-2*mod(pm,2); v0=v0+pm.*V0;

% Perturbation
XP1=1; V1=0.0; mode=1;
v0=v0+V1*sin(2*pi*x0/L*mode);
x0=x0+XP1*(L/N)*sin(2*pi*x0/L*mode);
out=(x0<0); x0(out)=x0(out)+L;
out=(x0>=L); x0(out)=x0(out)-L;

% calculate E0, satisfying the Gauss' Law
% solving the Poisson equation
p=1:N;p=[p p]; un=ones(NG-1,1);
Poisson=spdiags([un -2*un un],[-1 0 1],NG-1,NG-1);
g1=floor(x0/dx-.5)+1; g=[g1;g1+1];
fraz1=1-abs(x0/dx-g1+.5); fraz=[fraz1;1-fraz1];
out=(g<1);g(out)=g(out)+NG;
out=(g>NG);g(out)=g(out)-NG;
mat=sparse(p,g,fraz,N,NG);
rho=full((Q/dx)*sum(mat))'+rho_back;
Phi=Poisson\(-rho(1:NG-1)*dx^2);Phi=[Phi;0];
E0 =([Phi(NG); Phi(1:NG-1)]-[Phi(2:NG);Phi(1)])/(2*dx);

for it=1:NT
	% start computational cycle
	xkrylov = [v0; E0]; 
	% Newton Krylov GMRes solver
	[sol, it_hist, ierr] = nsolgm(xkrylov,'residueEC',tol); 
	v_average = sol(1:N);
	% update particle positions and velocities
	v0 = 2*v_average - v0;
	x0 = x0 + v_average*DT;
	% check if particle are out of the periodic boundaries
	out=(x0<0); x0(out)=x0(out)+L;	
	out=(x0>=L);x0(out)=x0(out)-L;
	% new electric field
	E0 = sol((N+1):(N + NG));
	% calculate the total energy
	Etot = 0.5*abs(Q)*sum(v0.^2) + 0.5*sum(E0.^2)*dx;
	% save the total energy
	histEnergy = [histEnergy Etot];
	% end computational cycle
end
\end{lstlisting}

\section{residueEC.m}
The Newton Krylov GMRes solver ({\em nsolgm.m}) requires the definition of a residue function ({\em residueEC.m}), where the discretized equations of the energy conserving PIC method are formulated. The following Matlab/Octave code presents the residue function for the energy conserving PIC code. The particle average velocity equations \ref{vnES} are defined at line 22, while the field equations \ref{MaxwDiscreteES} at line 30.
\begin{lstlisting}
% residual calculation for the EC PIC 
function res = residueEC(xkrylov)

global L; global dx; global NG; global DT; global N; global WP; 
global QM; global Q; global rho_back; 
global x0; global v0; global E0;

% calculate the x at n+1/2 time level
x_average = x0 + xkrylov(1:N)*DT/2;
% check if particle are out of the periodic boundaries
out=(x_average<0);x_average(out)=x_average(out)+L; 
out=(x_average>=L);x_average(out)=x_average(out)-L;
% interpolation
p=1:N;p=[p p]; g1=floor(x_average/dx-.5)+1; g=[g1;g1+1];
fraz1=1-abs(x_average(1:N)/dx-g1+.5); fraz=[(fraz1);1-fraz1];	
out=(g<1);g(out)=g(out) + NG; 
out=(g>NG);g(out)=g(out)- NG; 
mat=sparse(p,g,fraz,N,NG);

res = zeros(N + NG,1);
% average velocity residual 
res(1:N,1)=xkrylov(1:N)-v0-0.25*mat*QM*(E0+xkrylov((N+1):(N+NG)))*DT;

% calculate the average J
fraz=[(fraz1).*xkrylov(1:N);(1-fraz1).*xkrylov(1:N)];	
mat=sparse(p,g,fraz,N,NG);
J = full((Q/dx)*sum(mat))';

%  electric field residual
res((N+1):(N+NG))=xkrylov((N+1):(N+NG))-E0+J*DT;
\end{lstlisting}








\end{document}